\documentclass[unsortedaddress, pre, twocolumn, showpacs, amsmath, amssymb, superscriptaddress, flushbottom]{revtex4-1}

\usepackage{subfigure}
\usepackage{tikz}
\usepackage{bm}
\usepackage{amsmath}
\usepackage{amssymb}
\usepackage{tabularx}
\usepackage{bbm}
\usepackage[utf8]{inputenc}
\usepackage[T1]{fontenc}
\usepackage{graphicx}
\usepackage[english]{babel}
\usepackage{epsfig}
\usepackage{natbib}
\usepackage{times}
\usepackage{ae}

\newcommand{\del}{\partial}
\newcommand{\f}{\frac}

\newcommand{\be}{\begin{equation}}
\newcommand{\ee}{\end{equation}}
\newcommand{\ba}{\begin{array}}
\newcommand{\ea}{\end{array}}
\newcommand{\bc}{\begin{center}}
\newcommand{\ec}{\end{center}}

\begin{document}

\title{Efficient and accurate simulations of deformable particles immersed in a fluid using a combined immersed boundary lattice Boltzmann finite element method}

\date{\today}

\author{T.~Kr\"uger}
\email{t.krueger@mpie.de}
\affiliation{Max-Planck-Institut f\"ur Eisenforschung, Max-Planck-Str.~1, 40237~D\"usseldorf, Germany}
\author{F.~Varnik}
\affiliation{Interdisciplinary Center for Advanced Materials Simulation, Stiepeler Str.\ 129, 44780 Bochum, Germany}
\affiliation{Max-Planck-Institut f\"ur Eisenforschung, Max-Planck-Str.~1, 40237~D\"usseldorf, Germany}
\author{D.~Raabe}
\affiliation{Max-Planck-Institut f\"ur Eisenforschung, Max-Planck-Str.~1, 40237~D\"usseldorf, Germany}


\begin{abstract}
The deformation of an initially spherical capsule, freely suspended in simple shear flow, can be computed analytically in the limit of small deformations [D.\ Barth\'es-Biesel, J.\ M.\ Rallison, The Time-Dependent Deformation of a Capsule Freely Suspended in a Linear Shear Flow, J.\ Fluid Mech.\ 113 (1981) 251--267]. Those analytic approximations are used to study the influence of the mesh tessellation method, the spatial resolution, and the discrete delta function of the immersed boundary method on the numerical results obtained by a coupled immersed boundary lattice Boltzmann finite element method. For the description of the capsule membrane, a finite element method and the Skalak constitutive model [R.\ Skalak et al., Strain Energy Function of Red Blood Cell Membranes, Biophys.\ J.\ 13 (1973) 245--264] have been employed. Our primary goal is the investigation of the presented model for small resolutions to provide a sound basis for efficient but accurate simulations of multiple deformable particles immersed in a fluid. We come to the conclusion that details of the membrane mesh, as tessellation method and resolution, play only a minor role. The hydrodynamic resolution, i.e., the width of the discrete delta function, can significantly influence the accuracy of the simulations. The discretization of the delta function introduces an artificial length scale, which effectively changes the radius and the deformability of the capsule. We discuss possibilities of reducing the computing time of simulations of deformable objects immersed in a fluid while maintaining high accuracy.
\end{abstract}

\maketitle

%
%
%


\section{Introduction}
\label{sec:introduction}

Understanding the hydrodynamics of blood is certainly one of the major motivations for the simulation of deformable particles immersed in a fluid. Aside from the desire to investigate this aspect of fundamental research in more detail, there are many relevant applications in biology and medical sciences. The ultimate goal is the simulation of the human microcirculation up to the centimeter scale including the full dynamics of the cells, their interactions with each other and the blood vessel walls, and the impact of their microscopic properties on the macroscopic behavior of blood over the entire shear rate range. The complexity and scale-bridging of the coupled system of hydrodynamics and cell membrane dynamics requires numerical approaches to obtain meaningful results.

Capsules are elastic membranes filled with a fluid. The investigation of the dynamical response of capsules in fluids is not trivial since the hydrodynamic properties of the internal and external fluids, the constitutive model of the capsule membrane, and its shape affect the outcome. Fortunately, there are analytic solutions available for small deformations of single capsules in shear flow \cite{barths-biesel_motion_1980, barths-biesel_time-dependent_1981}. Experimental results for the behavior of artificial capsules have been obtained by \citet{chang_experimental_1993} and \citet{walter_shear-induced_2000}. \citet{pozrikidis_finite_1995} numerically studied those problems using the boundary element method (BEM). This method, in different formulations, has also been employed by, e.g., \citet{kraus_fluid_1996, ramanujan_deformation_1998, diaz_transient_2000, barths-biesel_effect_2002}, and \citet{lac_spherical_2004}. The BEM is only valid in Stokes flow and cannot be applied to flow situations with inertia. Fluid-capsule interactions using the immersed boundary method have been studied by \citet{eggleton_large_1998}. \citet{sui_hybrid_2008} have implemented a combined immersed boundary lattice Boltzmann method with grid refinement to study the transient deformation of capsules in simple shear flow at high resolutions. \citet{pozrikidis_effect_2001} considered bending resistance in capsule simulations. He states the importance of the presence of bending stiffness in biological cells. Simulations of multiple blood cells have been performed by, e.g., \citet{dupin_modelingflow_2007} and \citet{doddi_three-dimensional_2009}.

Although the processing power of present-day computers is large compared to that one or two decades ago, computing resources are still limited. In simulations of multiple deformable particles in flow, the spatial resolution must be kept sufficiently small in order to limit the computing time to a reasonable period. This calls for efficient numerical methods which are capable of capturing the dominant physical properties of the problem, even at small resolutions. Therefore, we aim at a better understanding of the behavior of the simulation technique employed in this paper, especially at smaller resolutions.

The lattice Boltzmann method (LBM) is a comparably new method to solve the full Navier-Stokes equations \cite{qian_lattice_1992, chen_lattice_1998, ladd2001lbs, succi_lattice_2001, sukop_lattice_2005}. The starting point is the lattice Boltzmann equation which is an approximate and discretized form of the Boltzmann equation. Virtual particles, also called populations, are moving on a lattice and collide at the fixed lattice nodes of the regular grid. In the macroscopic limit, the Navier-Stokes equations can be recovered from the well-defined collision rules of those particles. This method is particularly straightforward to implement, and it has proven to be accurate and applicable to many hydrodynamic problems, e.g., \cite{varnik_roughness-induced_2007, varnik_wetting_2008, ayodele_effect_2009, markus_acc}.

Peskin developed the immersed boundary method (IBM) \cite{peskin1972fpa, peskin_ibm_2002} to model blood flow in the heart. The strength of this method is that the immersed material is intrinsically deformable and that the Navier-Stokes equations do not need to be modified in the presence of the material, except for the inclusion of a body force. The IBM has especially attracted the attention of the scientists due to its capability to model thin and elastic membranes and therefore red blood cells and capsules in arbitrary external flow fields. In an attempt to benchmark and test the accuracy and convergence behavior of IBM, the IBM has been applied to stiff objects \cite{feng_immersed_2004, niu_momentum_2006, strack_three-dimensional_2007, peng_comparative_2008}. However, there are numerical limits to the applicability of IBM to stiff materials, and it requires some effort to do so. On the contrary, the lattice Boltzmann bounce-back (BB) scheme unrolls its benefits in the case of stiff obstacles in flow \cite{ginzbourg_local_1996, ladd2001lbs, lallemand_lattice_2003, ding_extension_2003, chun_interpolated_2007, peng_comparative_2008}, and it is much more demanding to capture deformable objects with BB.

In order to efficiently simulate deformable membranes immersed in a fluid, we use an approach combining the LBM as fluid solver, the IBM for the coupling of the fluid and the membranes, and a finite element method (FEM) for the computation of the membrane response to deformations (IBLBFEM). This approach has been employed successfully by other scientists, e.g., \citet{zhang_immersed_2007} or \citet{sui_hybrid_2008}. A similar method, with another Navier-Stokes solver, has also been applied by \citet{eggleton_large_1998} and \citet{doddi_three-dimensional_2009}. The advantage of this combined IBLBFEM is that the computations of the fluid and membranes are decoupled and that the meshes of the fluid and the membranes do not have to match. No remeshing is required for the membranes. The implementation is straightforward, and the method is powerful to simulate $O(100)$ deformable particles in flow solving the full Navier-Stokes equations and obtaining velocity, pressure, and shear stress information locally and at finite Reynolds numbers.

We use an explicit IBM coupling scheme which is efficient in terms of computing time. However, the no-slip condition at the membrane surface is not perfectly obeyed, and a drift of the capsules' volume can occur \cite{peskin_improved_1993, newren_unconditionally_2007}. For the simulation of stiff objects in flow (which is not the case here), modified IBM schemes exactly obeying the no-slip condition are known \cite{shu_novel_2007, wu_implicit_2009}. It is also often argued that the IBM is unstable in the limit of high stiffness \cite{tu_stability_1992}. There are methods to increase stability using either implicit or semi-implicit methods \cite{newren_unconditionally_2007, mori_implicit_2008}. Due to the softness of the capsules and the relatively short simulations in the present paper, instabilities do not occur, and the volume drift is negligible. We do not intent to comment on IBM-related stability issues or an improved implementation of the no-slip condition in this work.

When it comes to the discretization of the capsule, the question arises whether a structured or unstructured mesh should be used and how this mesh should be created. Structured meshes usually contain coordinate singularities whereas on unstructured meshes gradients have to be approximated. \citet{diaz_transient_2000} and \citet{lac_spherical_2004} have used structured meshes whereas \citet{kraus_fluid_1996, navot_elastic_1998}, and \citet{ramanujan_deformation_1998} have employed unstructured grids for the capsule. We use an unstructured mesh tessellation in this paper.

This paper is not targeted at finding new physics of capsules in shear flow. We rather wish to better understand the behavior of the combined IBLBFEM and the membrane tessellation on the accuracy and the numerical efficiency of the simulation of deformable capsules at small resolutions. This is the main difference between our recent effort and the work by \citet{sui_hybrid_2008} who have not discussed the behavior of their numerical method at small resolutions. Of special interest are the impact of the IBM interpolation stencil, the mesh tessellation, and the ratio between the average mesh node distance and the lattice constant of the LBM grid. We will also shortly discuss the importance of a well-chosen BGK LBM relaxation parameter. Those considerations are important for a correct setup of efficient simulations containing a large number of deformable objects with relatively coarse meshes. Therefore, our investigations are hoped to be useful for future simulations of multiple deformable particles in flow employing the IBM. We simulate the time evolution of the deformation of an initially spherical capsule freely suspended in an unbound simple shear flow. The interior and exterior fluids have the same properties, and they are Newtonian. The capsules have no bending resistance. For comparison, the approximated analytic steady-state solutions by \citet{barths-biesel_motion_1980} and \citet{barths-biesel_time-dependent_1981} are used.

The LBM will be shortly presented in Sec.\ \ref{subsec:lbm}, followed by an overview of the membrane model and the used FEM in Sec.\ \ref{subsec:membrane}. The IBM is briefly presented in Sec.\ \ref{subsec:ibm}, and the mesh influence is discussed in Sec.\ \ref{subsec:mesh}. In Sec.\ \ref{sec:theory}, the theory of small capsule deformations is shortly outlined. The simulations, results, and discussions can be found in Sec.\ \ref{sec:simulations}, followed by the conclusions in Sec.\ \ref{sec:conclusion}.


\section{Numerical methods}
\label{sec:methods}

The simulation algorithm consists of three major components: the fluid solver, the membrane model, and the coupling of fluid and membrane. For the fluid solver, we have used the D3Q19 Bhatnagar-Gross-Krook (BGK) lattice Boltzmann method (LBM). The membrane dynamics is derived from a constitutive model, and the strains in the capsule are evaluated using a finite element method (FEM). The interaction of the fluid and the membrane is captured by the immersed boundary method (IBM).

The overview of LBM is presented in Sec.\ \ref{subsec:lbm}, the membrane model is outlined in Sec.\ \ref{subsec:membrane}, and IBM is shortly covered in Sec.\ \ref{subsec:ibm}. Since also the discrete particle mesh plays a role in the simulations, its properties are briefly discussed in Sec.\ \ref{subsec:mesh}.

\subsection{Lattice Boltzmann method}
\label{subsec:lbm}

\begin{figure*}
\centering
\begin{tikzpicture}[z={(-0.45,-0.62)}]
\coordinate[z={(-0.45,-0.62)}] (v0) at (0,0,0);
\coordinate[z={(-0.45,-0.62)}] (v1) at (2,0,0);
\coordinate[z={(-0.45,-0.62)}] (v2) at (-2,0,0);
\coordinate[z={(-0.45,-0.62)}] (v3) at (0,2,0);
\coordinate[z={(-0.45,-0.62)}] (v4) at (0,-2,0);
\coordinate[z={(-0.45,-0.62)}] (v5) at (0,0,2);
\coordinate[z={(-0.45,-0.62)}] (v6) at (0,0,-2);
\coordinate[z={(-0.45,-0.62)}] (v7) at (2,0,2);
\coordinate[z={(-0.45,-0.62)}] (v8) at (-2,0,-2);
\coordinate[z={(-0.45,-0.62)}] (v9) at (-2,-2,0);
\coordinate[z={(-0.45,-0.62)}] (v10) at (2,2,0);
\coordinate[] (v11) at (-2,0,2);
\coordinate[] (v12) at (2,0,-2);
\coordinate[] (v13) at (2,-2,0);
\coordinate[] (v14) at (-2,2,0);
\coordinate[] (v15) at (0,-2,2);
\coordinate[] (v16) at (0,2,-2);
\coordinate[] (v17) at (0,-2,-2);
\coordinate[] (v18) at (0,2,2);
\draw[draw=gray!80,fill=gray!40,opacity=0.5] (v7)--(v12)--(v8)--(v11)--cycle;
\draw[draw=gray!80,fill=gray!40,opacity=0.5] (v15)--(v17)--(v16)--(v18)--cycle;
\draw[draw=gray!80,fill=gray!40,opacity=0.5] (v9)--(v13)--(v10)--(v14)--cycle;
\draw[-latex,very thick] (v0)--(v1) node[right]{1};
\draw[-latex,very thick] (v0)--(v2) node[left]{2};
\draw[-latex,very thick] (v0)--(v3) node[above]{3};
\draw[-latex,very thick] (v0)--(v4) node[below]{4};
\draw[-latex,very thick] (v0)--(v5) node[below left]{5};
\draw[-latex,very thick] (v0)--(v6) node[above right]{6};
\draw[-latex,thick,gray!90] (v0)--(v7) node[below]{7};
\draw[-latex,thick,gray!90] (v0)--(v8) node[above]{8};
\draw[-latex,thick,gray!90] (v0)--(v9) node[below left]{9};
\draw[-latex,thick,gray!90] (v0)--(v10) node[above right]{10};
\draw[-latex,thick,gray!90] (v0)--(v11) node[below left]{11};
\draw[-latex,thick,gray!90] (v0)--(v12) node[above right]{12};
\draw[-latex,thick,gray!90] (v0)--(v13) node[below right]{13};
\draw[-latex,thick,gray!90] (v0)--(v14) node[above left]{14};
\draw[-latex,thick,gray!90] (v0)--(v15) node[below]{15};
\draw[-latex,thick,gray!90] (v0)--(v16) node[above]{16};
\draw[-latex,thick,gray!90] (v0)--(v17) node[right]{17};
\draw[-latex,thick,gray!90] (v0)--(v18) node[left]{18};
\draw[z={(-0.45,-0.62)},-latex,very thick] (5,0,0)--(5,1,0) node[above]{y};
\draw[z={(-0.45,-0.62)},-latex,very thick] (5,0,0)--(5,0,1) node[below]{z};
\draw[z={(-0.45,-0.62)},-latex,very thick] (5,0,0)--(6,0,0) node[right]{x};
\end{tikzpicture}
\caption{Sketch of the D3Q19 lattice structure. All velocity vectors are located in at least one of the three coordinate planes (light gray). The velocity vectors either point to the next neighbors along the coordinate axes ($\bm c_{1-6}$, black arrows) or to the next but one neighbors ($\bm c_{7-18}$, dark-gray arrows). The zero velocity $\bm c_0$ is not shown.}
\label{fig:d3q19}
\end{figure*}
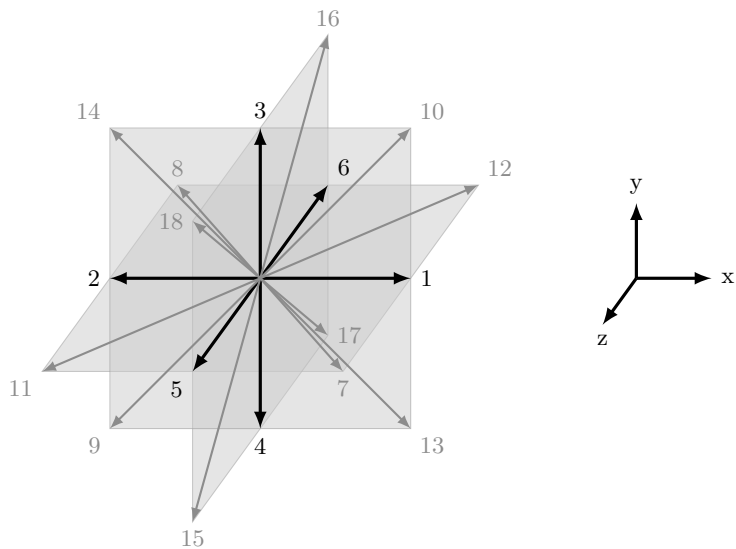

In the last two decades, the LBM has become a competitive Navier-Stokes solver with increasing prominence among scientists in the field of computational fluid dynamics \cite{qian_lattice_1992, chen_lattice_1998, ladd2001lbs, guo_discrete_2002, kruger_shear_2009}. Its strength is based on its simple coding and, since LBM is an automaton, its locality, making it intrinsically parallelizable.

The lattice Boltzmann equation (LBE), Eq.\ (\ref{eq:lbequation}), can be regarded as a discretized form of the Boltzmann equation. On the other hand, it is an extension of the lattice gas cellular automaton \cite{frisch_lattice-gas_1986}. In contrast to conventional Navier-Stokes solvers, the LBE is not the discretized form of the Navier-Stokes equations. While conventional methods directly solve the Navier-Stokes equations in terms of the pressure $p$ and the velocity $\bm u$, the LBM introduces a number of $q$ populations $f_i$ ($i = 0, \ldots, q - 1$) streaming along a regular lattice (lattice constant $\Delta x$) in discrete time steps. Those populations can be regarded as mesoscopic particle packets propagating and colliding.

The evolution of the populations $f_i$ is given by the LBE, which takes the form
\begin{widetext}
\begin{equation}
\label{eq:lbequation}
f_i (\bm x + \bm c_i \Delta t, t + \Delta t) - f_i (\bm x, t) = - \f{1}{\tau} \left(f_i (\bm x, t) - f_i^{\text{eq}} (\bm x, t)\right) + F_i \Delta t
\end{equation}
\end{widetext}
in the BGK approximation. The dimensionless relaxation parameter $\tau$ of the fluid is connected to the speed of sound $c_s$ and the kinematic viscosity $\nu$ by $\nu = c_s^2 (\tau - 1/2) \Delta t$, where $c_s = \sqrt{1/3}\, \Delta x / \Delta t$ holds. At each time step $t$, the populations propagate along the $q$ discretized velocity vectors $\bm c_i$ to the next neighbors. At those points, they collide according to the right-hand side of Eq.\ (\ref{eq:lbequation}). The significance of $F_i$ is explained below. The equilibrium populations are given by
\be
\label{eq:equilibrium}
f_i^{\text{eq}} = w_i\, \rho \left(1 + 3 \bm c_i \cdot \bm u + \f 92 (\bm c_i \cdot \bm u)^2 - \f 32 \bm u \cdot \bm u\right).
\ee
This is closely related to the truncated form of the Maxwell distribution which is a very good approximation for small Mach numbers. The $q$ factors $w_i$ are the lattice weights, depending on the underlying lattice structure. Their choice ensures the isotropy of the fluid, a necessity to solve the Navier-Stokes equations asymptotically. In the present paper, we use a 3D model with $19$ velocities, designated D3Q19. The lattice structure, the corresponding velocities $\bm c_i$, and the lattice weights $w_i$ are introduced in \cite{qian_lattice_1992}. A sketch of the D3Q19 lattice is shown in Fig.\ \ref{fig:d3q19}.

A body force density $\bm f$ can be incorporated via $F_i$ in Eq.\ (\ref{eq:lbequation}) \cite{ladd2001lbs, guo_discrete_2002},
\be
\label{eq:latticeforce}
F_i = \left(1 - \f{1}{2 \tau}\right) w_i \left(\f{\bm c_i - \bm u}{c_s^2} + \f{\bm c_i \cdot \bm u}{c_s^4} \bm c_i\right) \cdot \bm f.
\ee
This force density is particularly important for the coupling of the fluid and the immersed membranes, but it is also commonly used to include gravity. More details are given in Sec.\ \ref{subsec:ibm}.

Finally, the macroscopic properties of the fluid have to be extracted from the populations $f_i$. The density and the velocity can directly be recovered by computing the zeroth and first moments:
\begin{align}
\label{eq:lbmmomenta0}
\rho &= \sum_i f_i, \\
\label{eq:lbmmomenta1}
\rho\, \bm u &= \sum_i \bm c_i f_i + \f{\Delta t}{2} \bm f,
\end{align}
at each fluid lattice node. The deviatoric shear stress tensor $\bm \sigma$ can also be computed from the populations locally \cite{kruger_shear_2009}.

The simple shear flow required for the present simulations can be realized by using the bounce-back method for moving walls \cite{ladd2001lbs}. The fluid is fully periodic along the $x$- and $y$-axes (velocity and vorticity directions, respectively), but it is bound by two plane walls at $z = \pm H/2$, where $H$ is the distance of the walls. Moving the walls in the $x$-direction with velocity $\pm u_w$ ($u_w > 0$ for $z > 0$), the shear rate of the fluid is $\dot \gamma = 2 u_w / H$.

In order to obtain accurate predictions, the limits of the validity of the LBM must be acknowledged. The slip velocities $u_w$ have to be chosen sufficiently small so that the LBM operates in the small Mach number limit. An even more stringent restriction of the wall speed is given by the Reynolds number. The theory of small deformations of a capsule in simple shear flow is only valid in Stokes flow, $\text{Re} \ll 1$. The correct choice of the relaxation parameter $\tau$ also influences the accuracy of the simulations \cite{holdych_truncation_2004, kruger_shear_2009}, especially in combination with the IBM \cite{le_boundary_2009}. The LBE intrinsically contains the partial time derivative $\partial \bm u / \partial t$ of the Navier-Stokes equations. Hence, Stokes flow can only be reached asymptotically.

A more detailed presentation of the LBM can be found in the literature, e.g., in the monographs by \citet{succi_lattice_2001} or \citet{sukop_lattice_2005}.

\subsection{Membrane model and force computation}
\label{subsec:membrane}

\begin{figure*}
\centering
\begin{tikzpicture}[]
\node at (1.5,-0.5) {(a)};
\coordinate (u1) at (0,0);
\coordinate (u2) at (2.8,1.1);
\coordinate (u3) at (0.6,2.95);
\node[left] at (u1) {$1$};
\node[right] at (u2) {$2$};
\node[above] at (u3) {$3$};
\draw[draw=white] (u1)--(u2) node [midway, below right] {$l'_0$};
\draw[draw=white] (u1)--coordinate [near start](anker)(u3) node [midway, above left] {$l_0$};
\draw[draw=white] (u3)--(u2);
\draw[fill=lightgray,thick,semitransparent] (u1)--(u2)--(u3)--cycle;
\draw[thick] (anker) arc (70:28:1cm);
\node at (0.7,0.8) {$\varphi_0$};
\node at (4.5,-0.5) {(b)};
\coordinate (d1) at (3.0,2.6);
\coordinate (d2) at (6.0,0.2);
\coordinate (d3) at (4.6,3.1);
\node[left] at (d1) {$1$};
\node[below] at (d2) {$2$};
\node[above] at (d3) {$3$};
\draw[draw=white] (d1)--(d2) node [midway, below left] {$l'$};
\draw[draw=white] (d1)--coordinate [pos=0.4](anker2)(d3) node [midway, above left] {$l$};
\draw[draw=white] (d3)--(d2);
\draw[fill=lightgray,thick,semitransparent] (d1)--(d2)--(d3)--cycle;
\draw[thick] (anker2) arc (15:-24:1cm);
\node at (3.9,2.4) {$\varphi$};
\node at (8.5,-0.5) {(c)};
\coordinate (n1) at (7,0);
\coordinate (n2) at (10,0);
\coordinate (n3) at (8.5,2.6);
\coordinate (n4) at (10.6,0);
\coordinate (n5) at (7.5,1.5);
\draw[fill=lightgray,thick,semitransparent] (n1)--(n2)--(n3)--cycle;
\draw[fill=lightgray,thick,semitransparent] (n1)--(n4)--(n5)--cycle;
\draw[thick,-latex] (n1)--++(0.7,0) node[midway,below] {$x$};
\draw[thick,-latex] (n1)--++(0,0.7) node[midway,left] {$y$};
\draw[-latex,thick] (n3)--(n5) node [midway,above left] {$\bm v_3$};
\draw[-latex,thick] (n2)--(n4) node [midway,below] {$\bm v_2$};
\end{tikzpicture}
\caption{Illustration of (a) the equilibrium face (defined by $l_0$, $l'_0$, and $\varphi_0$), (b) its deformed shape (accordingly defined by $l$, $l'$, and $\varphi$), and (c) both transformed to the same $xy$-plane. The displacement vector $\bm v_1$ is identically zero, and the other two are shown in subfigure (c). The deformation state $(\lambda_1, \lambda_2)$ of the face is then uniquely defined.}
\label{fig:deformedface}
\end{figure*}
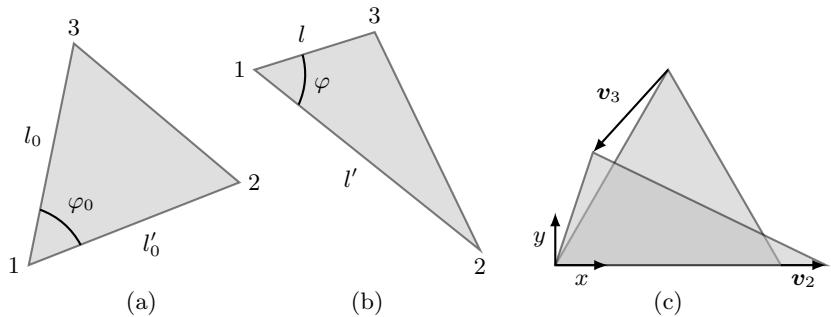

The IBM algorithm requires knowledge of the forces at the nodes of the tessellated membrane (cf.\ Sec.\ \ref{subsec:ibm}). For a hyperelastic material, i.e., negligible viscous and plastic forces, the shear forces can be computed from a constitutive model for the areal strain energy density $w^S$. The contributions to the total energy $W$ can generally be written in the form $W = W^S + W^B + W^A + W^V$, where the superscripts denote strain, bending, surface, and volume contributions, respectively.

The areal strain energy density $w^S$ obeying $W^S = \int \text{d}A\, w^S$ ($\text{d}A$ is the surface element) can only depend on the invariants $I_1 = \lambda_1^2 + \lambda_2^2 - 2$ and $I_2 = \lambda_1^2 \lambda_2^2 - 1$ for a thin membrane with isotropic and homogeneous elastic properties, i.e., per definition $w^S$ is invariant under rotations and translations. $\lambda_1$ and $\lambda_2$ are the local principal in-plane stretch ratios. Deformations of biological cells can be large, and thus the linear strain-stress approximation is not justified in general. \citet{skalak_strain_1973} have suggested an energy model which is able to reproduce experimental data of red blood cells at both small and large strains,
\be
\label{eq:constitutivelaw}
w^S = \f{k_s}{12} \left(I_1^2 + 2 I_1 - 2 I_2\right) + \f{k_\alpha}{12} I_2^2.
\ee
The surface elastic shear modulus $k_s$ and area dilation modulus $k_\alpha$ control the strength of the membrane response to deformation (shear and dilation). A commonly used model is the neo-Hookean law which is equivalent to the zero-thickness shell membrane proposed by \citet{ramanujan_deformation_1998} for small deformations. Another constitutive model has been proposed by \citet{navot_elastic_1998}. More information about those constitutive laws can be found in the literature, e.g., \cite{barths-biesel_effect_2002, sui_hybrid_2008} and will not be discussed here. For all our simulations, we have employed the Skalak membrane model, Eq.\ (\ref{eq:constitutivelaw}).

The bending energy $W^B$ can have local and non-local contributions. We have not considered any bending energy in the present simulations, i.e., $W^B = 0$ since it is not considered in the analytic investigation by \citet{barths-biesel_time-dependent_1981}. However, it has been thoroughly discussed in the literature that a bending resistance has to be included whenever strong local curvatures appear. Elsewise, the membranes can buckle or collapse \cite{eggleton_large_1998, ramanujan_deformation_1998, pozrikidis_effect_2001, bagchi_computational_2005}. This is especially the case for more complex geometries and strong deformations as in the case of red blood cells. More details about the form of the bending energy are provided in \citet{canham_minimum_1970, helfrich_elastic_1973, svetina_membrane_1989}, and \citet{gompper2008soft}.

The total volume and surface of the membrane may be restricted. This can be formulated by defining global volume and surface energies, $W^V$ and $W^A$. Those energies are minimum if the volume and surface equal their corresponding equilibrium values \cite{gompper2008soft}. In the present simulations, we have neither employed a volume nor a surface energy, i.e., $W^V = W^A = 0$. Although the IBM is not perfectly volume-conserving, cf.\ Sec.\ \ref{subsec:ibm}, the volume drift is almost negligible in most of the present simulations of a single capsule. The reason is the relatively short duration of the simulations. We stress that volume and surface energies may have to be considered to improve numerical stability in longer simulations with complex flow fields, large local shear rates, and coarse mesh resolutions.

The capsule membrane is numerically described by a number $N_f$ of flat triangular face elements, which remain flat even at large deformations. While the deformation state is a property of the faces, the membrane forces have to be known at the corners of the faces (nodes). The first step in the computation of the strains $\lambda_{1,2}$ of a given face element is the identification of the displacements $\bm v_i$ ($i = 1, 2, 3$) of the nodes. The deformed and undeformed elements are transformed to a common plane (here: $xy$-plane) in such a way that the edges $l'_0$ and $l'$ are aligned, cf.\ Fig.\ \ref{fig:deformedface}. There is no restriction since translations and rotations do not change the energy of an element. The basic assumption is that the displacement gradient tensor $(D_{\alpha \beta}) \approx (\delta_{\alpha \beta} + \partial_\beta v_\alpha)$ is spatially constant over the entire face element. This can be realized by introducing a linear shape function $N_i (x, y) = a_i x + b_i y + c_i$ ($i = 1, 2, 3$) for each node. The coefficients are found by letting $N_i(x_j, y_j) = \delta_{ij}$ ($i, j = 1, 2, 3$), i.e., each shape function $N_i$ is unity at the location of the corresponding node $i$, but zero at the two nodes other than $i$. The linear displacement field of the face element can then be written as
\be
\bm v(x, y) = N_1 \bm v_1 + N_2 \bm v_2 + N_3 \bm v_3,
\ee
and the displacement gradient tensor $\bm D$ can be computed. Its components do not depend on $x$ or $y$, but on the shape function coefficients $a_i$ and $b_i$ which are fixed by the shape of the undeformed element only, i.e., $a_i$ and $b_i$ are constant in time for each node in the face. It is straightforward to show that the displacement gradient tensor then has the form $(D_{\alpha \beta}) = \left(\begin{smallmatrix} a & b \\ 0 & c\end{smallmatrix}\right)$ with \cite{gompper2008soft}
\begin{align}
a &= \f{l}{l_0}, \\
b &= \f{1}{\sin \varphi_0} \left(\f{l'}{l'_0} \cos \varphi - \f{l}{l_0} \cos \varphi_0\right), \\
\label{eq:c}
c &= \f{l'}{l'_0} \f{\sin \varphi}{\sin \varphi_0}.
\end{align}
Here, $l$ and $l'$ are the lengths of two arbitrary edges of the face, and $\varphi$ is the angle between those edges, cf.\ Fig.\ \ref{fig:deformedface}. The zero index ($l_0$, $l'_0$, and $\varphi_0$) denotes the undeformed values. The current deformation of a face is evaluated from the equations 
\begin{align}
\label{eq:lambda1}
\lambda_1^2 \lambda_2^2 &= a^2 c^2, \\
\lambda_1^2 + \lambda_2^2 &= a^2 + b^2 + c^2
\end{align}
since $\lambda_1^2 + \lambda_2^2 = \text{tr}\, \bm D^{\text{T}} \bm D$ and $\lambda_1^2 \lambda_2^2 = \text{det}\, \bm D^{\text{T}} \bm D$. Note that the product $\bm D^{\text{T}} \bm D$ is rotationally invariant. For more details, we refer to \citet{charrier_free_1989, shrivastava_large_1993}, and \citet{gompper2008soft}.

The total energy in the present simulations has shear contributions only, $W = W^S$. The strain energy $W^S$ is computed from the areal energy density $w^S$, Eq.\ (\ref{eq:constitutivelaw}), and the local reference area $A_0$ of the membrane face elements \cite{charrier_free_1989, shrivastava_large_1993, gompper2008soft},
\be
W^S = \sum_j^{\text{faces}} A_{0j} w^S_j.
\ee
Once the energy of the membrane is known, the forces acting on the fluid exerted by node $i$ at position $\bm x_i$ can be computed from the principle of virtual work,
\be
\label{eq:forcefromenergy}
\bm F_i = - \f{\del W(\bm x_i)}{\del \bm x_i}.
\ee
This procedure is equivalent to the approach explained in details in \cite{charrier_free_1989, shrivastava_large_1993}.

\subsection{Immersed boundary method}
\label{subsec:ibm}

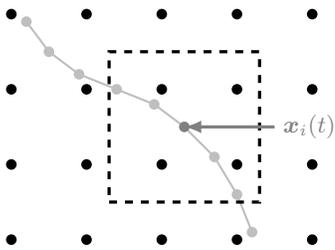
\begin{figure}
\centering
\begin{tikzpicture}[]
\coordinate (lu) at (1.3,0.5);
\coordinate (lo) at (1.3,2.5);
\coordinate (ru) at (3.3,0.5);
\coordinate (ro) at (3.3,2.5);
\fill (0,0) circle (2pt) node[]{};
\fill (0,1) circle (2pt) node[]{};
\fill (0,2) circle (2pt) node[]{};
\fill (0,3) circle (2pt) node[]{};
\fill (1,0) circle (2pt) node[]{};
\fill (1,1) circle (2pt) node[]{};
\fill (1,2) circle (2pt) node[]{};
\fill (1,3) circle (2pt) node[]{};
\fill (2,0) circle (2pt) node[]{};
\fill (2,1) circle (2pt) node[]{};
\fill (2,2) circle (2pt) node[]{};
\fill (2,3) circle (2pt) node[]{};
\fill (3,0) circle (2pt) node[]{};
\fill (3,1) circle (2pt) node[]{};
\fill (3,2) circle (2pt) node[]{};
\fill (3,3) circle (2pt) node[]{};
\fill (4,0) circle (2pt) node[]{};
\fill (4,1) circle (2pt) node[]{};
\fill (4,2) circle (2pt) node[]{};
\fill (4,3) circle (2pt) node[]{};
\coordinate (mem1) at (0.2,2.9);
\coordinate (mem2) at (0.5,2.5);
\coordinate (mem3) at (0.9,2.2);
\coordinate (mem4) at (1.4,2.0);
\coordinate (mem5) at (1.9,1.8);
\coordinate (mem6) at (2.3,1.5);
\coordinate (mem7) at (2.7,1.1);
\coordinate (mem8) at (3.0,0.6);
\coordinate (mem9) at (3.2,0.1);
\draw[lightgray,thick] (mem1)--(mem2)--(mem3)--(mem4)--(mem5)--(mem6)--(mem7)--(mem8)--(mem9);
\fill[lightgray] (mem1) circle (2pt) node[]{};
\fill[lightgray] (mem2) circle (2pt) node[]{};
\fill[lightgray] (mem3) circle (2pt) node[]{};
\fill[lightgray] (mem4) circle (2pt) node[]{};
\fill[lightgray] (mem5) circle (2pt) node[]{};
\fill[gray] (mem6) circle (2pt) node[]{};
\fill[lightgray] (mem7) circle (2pt) node[]{};
\fill[lightgray] (mem8) circle (2pt) node[]{};
\fill[lightgray] (mem9) circle (2pt) node[]{};
\draw[dashed,very thick] (lu)--(lo)--(ro)--(ru)--cycle;
\coordinate (text) at (3.5,1.5);
\node[gray,right] at (text) {$\bm x_i(t)$};
\draw[-latex,gray,very thick] (text)--(mem6);
\end{tikzpicture}
\caption{Illustration of the immersed boundary method in two dimensions. The membrane mesh (light gray) moves on top of the fixed fluid lattice (black). If the 2-point interpolation stencil is used, only the four lattice nodes enclosed by the dashed square with side length $2 \Delta x$ are required for the spreading and interpolation steps of membrane node $i$ located at $\bm x_i(t)$ (dark-gray).}
\label{fig:ibm}
\end{figure}

The IBM was originally proposed by Peskin \cite{peskin1972fpa, peskin_ibm_2002}. The basic idea is to couple the Eulerian coordinate system of the fluid lattice and the arbitrary Lagrangian coordinate system of a surface which is not conform to the regular lattice. The IBM is a front-tracking coupling method.

Caused by its deformations, the membrane exerts a force $\bm F_i(t)$ on the fluid at time step $t$. The fluid lattice nodes have fixed positions $\bm X$ and the membrane nodes $i$ are located at $\bm x_i(t)$. In the discretized description, the Eulerian body force density $\bm f(\bm X, t)$ is computed from the Lagrangian force $\bm F_i(t)$ by the spreading operation
\be
\label{eq:forcespreading}
\bm f(\bm X, t) = \sum_i \bm F_i(t)\, \delta(\bm X - \bm x_i(t)).
\ee
The lattice force density $\bm f(\bm X, t)$ is then used in the lattice Boltzmann equation, Eq.\ (\ref{eq:lbequation}), via the force coupling, Eq.\ (\ref{eq:latticeforce}). The kernel $\delta(\bm X - \bm x_i(t))$ is a discretized Dirac delta function with a finite support. \citet{peskin_ibm_2002} has shown that this function has to obey some basic properties to maintain momentum and angular momentum conservation. Still, the width of the function is not restricted a priori, and it can be considered as a free parameter of the IBM. We use the common decomposition $\delta(\bm r) = \phi(x) \phi(y) \phi(z)$. Among others \cite{peskin_ibm_2002, yang_smoothing_2009}, the most popular interpolation functions $\phi_n$ with a support of $n = 2, 3, 4$ lattice nodes along each coordinate axis are
\begin{widetext}
\begin{align}
\label{eq:2point}
\phi_2(r) &= \begin{cases} 1 - |r| & 0 \leq |r| \leq 1 \\ 0 & 1 \leq |r|\end{cases}, \\
\label{eq:3point}
\phi_3(r) &= \begin{cases} \f13 \left(1 + \sqrt{1 - 3 r^2}\right) & 0 \leq |r| \leq \f12 \\ \f16 \left(5 - 3 |r| - \sqrt{-2 + 6 |r| - 3 r^2}\right) & \f12 \leq |r| \leq \f32 \\ 0 & \f32 \leq |r|\end{cases}, \\
\label{eq:4point}
\phi_4(r) &= \begin{cases}\f18\left(3 - 2 |r| + \sqrt{1 + 4 |r| - 4 r^2}\right) & 0 \leq |r| \leq 1 \\ \f18 \left(5 - 2 |r| - \sqrt{-7 + 12 |r| - 4 r^2}\right) & 1 \leq |r| \leq 2 \\ 0 & 2 \leq |r|\end{cases}.
\end{align}
\end{widetext}
Completing the coupling of the fluid and the membrane, the new velocities $\bm u_i(t+1)$ of the membrane nodes $i$ are computed in the interpolation step, using the new lattice velocities but the old node positions,
\be
\label{eq:velocityinterpolation}
\bm u_i(t + \Delta t) = \sum_{\bm X} \bm u(\bm X, t + \Delta t)\, \delta(\bm X - \bm x_i(t)).
\ee
Here, the no-slip condition is assumed to be valid at the location of the membrane, i.e., the membrane moves with the ambient fluid velocity. The interpolation functions in Eqs.\ (\ref{eq:forcespreading}) and (\ref{eq:velocityinterpolation}) are the same. The principle of the IBM is illustrated in Fig.\ \ref{fig:ibm}. Finally, the membrane nodes $i$ are advected explicitly by the Euler rule
\be
\label{eq:positionupdate}
\bm x_i(t + \Delta t) = \bm x_i(t) + \bm u_i(t + \Delta t) \Delta t.
\ee
We have found that the Adams-Bashforth scheme
\be
\bm x_i(t + \Delta t) = \bm x_i(t) + \left(\f 32 \bm u_i(t + \Delta t) - \f 12 \bm u_i(t)\right) \Delta t
\ee
which has been used by \citet{doddi_lateral_2008} does not change the results for simulations of short duration. Yet, it provides additional accuracy for long-time simulations since it is a second-order scheme.

Although the no-slip condition at the membrane surface can be exactly fulfilled in the continuous limit, this is not the case in the discretized, explicit version of IBM. Even for an incompressible velocity field, the standard interpolation algorithm shown in this section does not assure that the volume of a closed membrane remains exactly constant in time. This problem has been recognized early, and improved immersed boundary approaches have been proposed for example by \citet{peskin_improved_1993} and \citet{wu_simulation_2009}. Due to the comparably short simulation times in the present paper, there is no need to counteract the volume drift.

In conclusion, each time step of the combined IBLBFEM scheme consists of the following sub-steps (we set $\Delta t = 1$).
\begin{enumerate}
\item At the beginning of time step $t$, the membrane node positions $\bm x_i(t)$ and the entire fluid state $\bm u(\bm X, t)$, $\rho(\bm X, t)$ are known. From the displacements of the membrane nodes, the forces $\bm F_i(t)$ are computed using the FEM (Sec.\ \ref{subsec:membrane}).
\item The membrane forces $\bm F_i(t)$ are spread to the Eulerian grid via IBM, Eq.\ (\ref{eq:forcespreading}), and the body force density $\bm f(\bm X, t)$ is obtained.
\item $\bm f(\bm X, t)$ is used in the LBM to compute the new state of the fluid, $\bm u(\bm X, t+1)$, $\rho(\bm X, t+1)$ (Sec.\ \ref{subsec:lbm}).
\item The new velocities $\bm u_i(t+1)$ of the membrane nodes are computed in the framework of IBM, Eq.\ (\ref{eq:velocityinterpolation}).
\item The new positions of the membrane nodes $\bm x_i(t+1)$ are found by evaluating Eq.\ (\ref{eq:positionupdate}).
\item Information on the membrane and fluid state \textit{after} time step $t$ may be written to the disk, using $\bm x_i(t+1)$, $\bm u(\bm X, t+1)$, etc. Get back to the first sub-step and recompute for time step $t+1$.
\end{enumerate}

\subsection{Membrane tessellation}
\label{subsec:mesh}

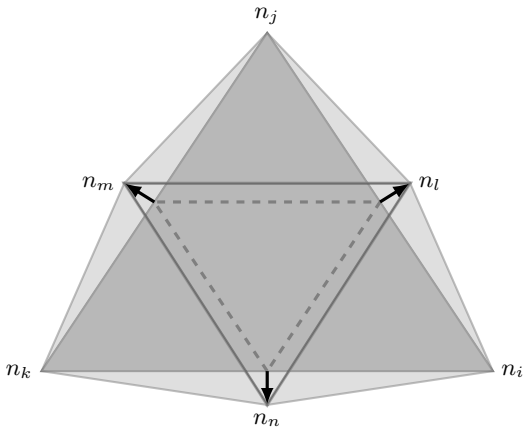
\begin{figure}
\centering
\begin{tikzpicture}[]
\coordinate (nk) at (0,0);
\coordinate (ni) at (6,0);
\coordinate (nj) at (3,4.5);
\draw[] (nk)node[left]{$n_k$}--(ni)node[right]{$n_i$}--(nj)node[above]{$n_j$}--cycle;
\draw[draw=gray,fill=gray!60,thick] (nk)--(ni)--(nj)--cycle;
\coordinate (nn_old) at (3,0);
\coordinate (nl_old) at (4.5,2.25);
\coordinate (nm_old) at (1.5,2.25);
\draw[draw=darkgray,dashed,very thick] (nn_old)--(nm_old)--(nl_old)--cycle;
\coordinate (nn) at (3,-0.45);
\coordinate (nl) at (4.9,2.5);
\coordinate (nm) at (1.1,2.5);
\draw[] (nn)node[below]{$n_n$}--(nl)node[right]{$n_l$}--(nm)node[left]{$n_m$}--cycle;
\draw[draw=darkgray,very thick,fill=lightgray,semitransparent] (nn)--(nm)--(nl)--cycle;
\draw[-latex,very thick] (nn_old)--(nn);
\draw[-latex,very thick] (nl_old)--(nl);
\draw[-latex,very thick] (nm_old)--(nm);
\draw[draw=gray,fill=gray!50,semitransparent,thick] (nn)--(nk)--(nm)--cycle;
\draw[draw=gray,fill=gray!50,semitransparent,thick] (nm)--(nj)--(nl)--cycle;
\draw[draw=gray,fill=gray!50,semitransparent,thick] (nl)--(ni)--(nn)--cycle;
\end{tikzpicture}
\caption{Illustration of the mesh subdivision. The face defined by the nodes $n_i$, $n_j$, and $n_k$ (dark-gray) is subdivided. First, the edges are halved and new nodes $n_l$, $n_m$, $n_n$ are created. The six nodes are connected in such a way that four faces of equal area are produced (dashed lines). Finally, the nodes $n_l$, $n_m$, and $n_n$ are radially shifted, until they are located on the sphere enclosing the body. The four final faces are shown in light gray.}
\label{fig:subdivision}
\end{figure}

\begin{figure*}
\centering
\subfigure[\label{fig:CGAL} CGAL]{\includegraphics[width=0.3\linewidth,clip=true]{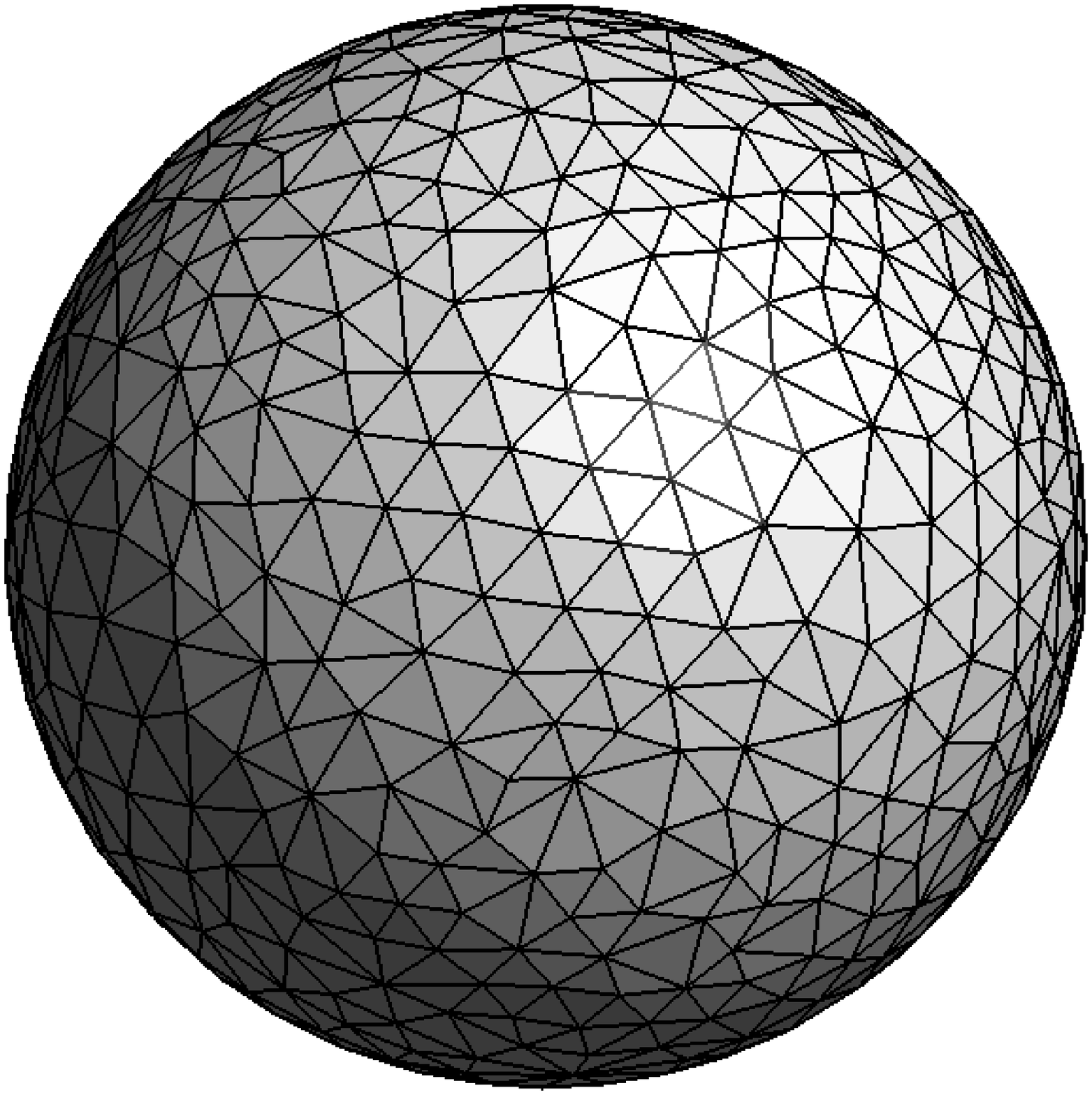}} \hfill
\subfigure[\label{fig:Gmsh} Gmsh]{\includegraphics[width=0.3\linewidth,clip=true]{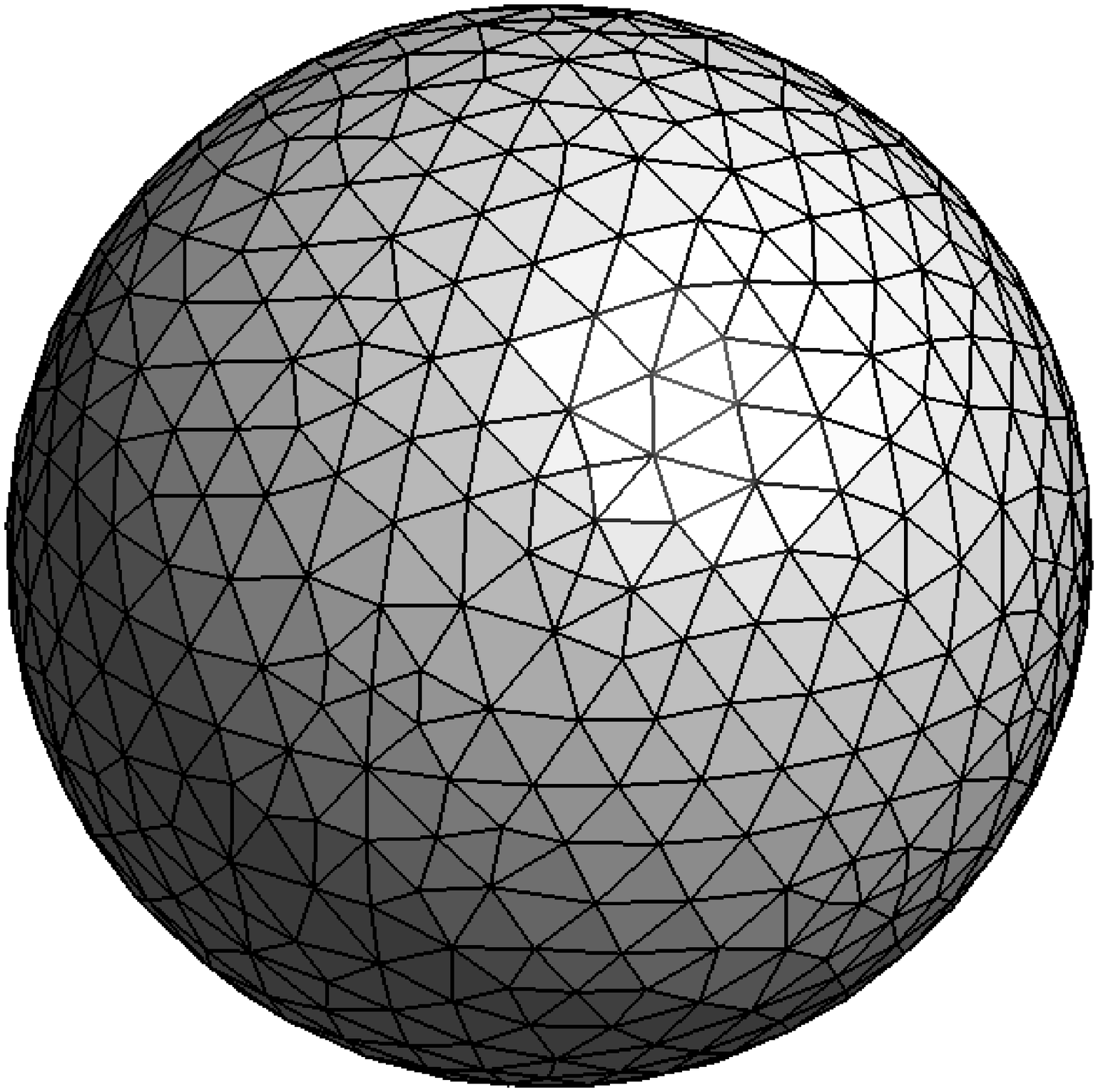}} \hfill
\subfigure[\label{fig:ico} icosahedron]{\includegraphics[width=0.3\linewidth,clip=true]{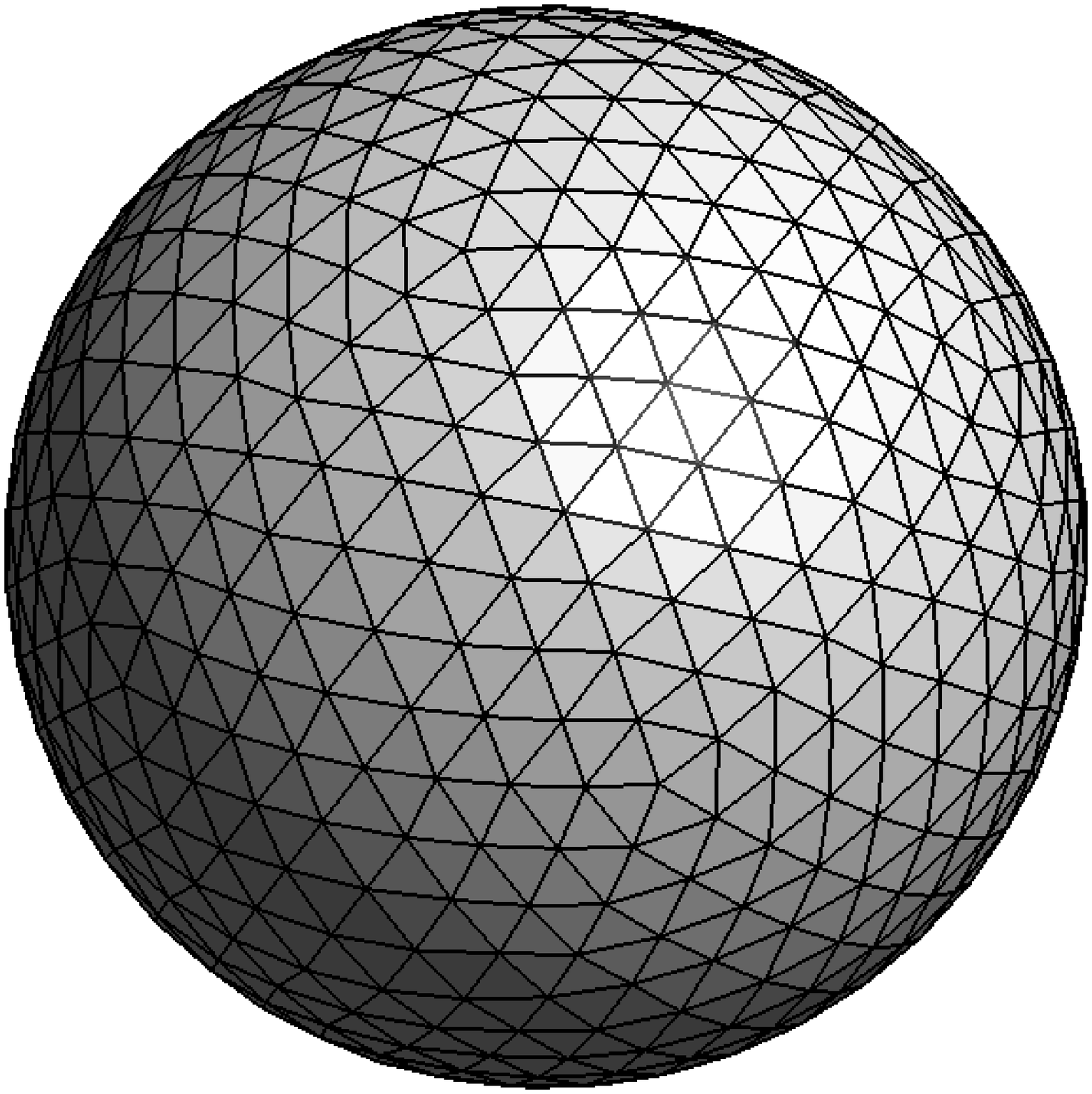}}
\caption{Meshes produced by \subref{fig:CGAL} CGAL, \subref{fig:Gmsh} Gmsh, and \subref{fig:ico} successive icosahedron subdivision. The meshes have $N_f = 1278$, $1284$, and $1280$ faces, respectively. Mesh \subref{fig:ico} has superior isotropy and homogeneity properties.}
\label{fig:meshtypes}
\end{figure*}

\begin{table*}
\begin{ruledtabular}
\centering
\caption{\label{tab:mesh_quality} Properties of spherical meshes, created with CGAL, Gmsh, and a successive subdivision of a regular icosahedron. The spheres have $N_f \approx 1280$ nodes each. $N_{nn}^<$ and $N_{nn}^>$ are the smallest and largest number of neighboring nodes to any node found for the given mesh. $A$ is the face area, $l$ is the edge length, $\varphi_n$ is the angle between neighboring face normals, and $\varphi_e$ is the angle between neighboring edges (both being members of a common face). The bar indicates the average of a quantity taken over the entire mesh, $\sigma$ denotes the standard deviation. Clearly, the mesh based on the icosahedron has superior quality in terms of isotropy and homogeneity.}
\begin{tabularx}{\linewidth}{crrr}
quantity & CGAL & Gmsh & subdivision \\ \hline
$N_f$ & $1278$ & $1284$ & $1280$ \\
$N_{nn}^<$, $N_{nn}^>$ & 4, 10 & 4, 8 & 5, 6 \\
$\sigma_{A} / \bar A$ & $25.8\%$ & $21.2\%$ & $8.6\%$ \\
$\sigma_{l} / \bar l$ & $19.0\%$ & $14.3\%$ & $6.5\%$ \\
$\sigma_{\varphi_n} / \bar \varphi_n$ & $42.5\%$ & $26.2\%$ & $15.9\%$ \\
$\sigma_{\varphi_e} / \bar \varphi_e$ & $25.1\%$ & $17.2\%$ & $9.3\%$
\end{tabularx}
\end{ruledtabular}
\end{table*}

The question arises whether the detailed properties of the mesh tessellation have a significant impact on the quality of the simulation results or not. There are different approaches for the generation of a spherical mesh. We will focus on three of them:
\begin{enumerate}
 \item tessellation of an implicit surface using the CGAL libraries \cite{cgal:ry-smg-08},
 \item finite element mesh generation using Gmsh \cite{Gmsh:website},
 \item successive subdivision, starting from a coarse mesh of high symmetry.
\end{enumerate}

The CGAL libraries \cite{cgal:ry-smg-08} allow the user to tessellate the surface defined by the zero level set of any implicit function $F(x, y, z)$. For a sphere with radius $r$, this function reads $F(x, y, z) = x^2 + y^2 + z^2 - r^2$. One has control over the number of faces and the distance between neighboring nodes, but the mesh suffers from a reduced homogeneity and isotropy. The advantage of this method is that any implicit surface can be tessellated without much effort.

Gmsh \cite{Gmsh:website} is not able to tessellate implicit surfaces, but the user can construct geometric objects like spheres. The surface of those shapes can then be tessellated. Also here, the high isotropy of the initial sphere is not completely captured by the mesh.

A method to produce a spherical mesh of high homogeneity and isotropy is subdivision of a highly symmetric mesh of low resolution. We have used a regular icosahedron, one of the five Platonic solids. It has 20 equilateral triangles as faces, 12 nodes and 30 edges of equal length. The numbers of nodes $N_n$ and faces $N_f$ of any closed surface consisting only of triangles are related by $2 N_n = N_f + 4$. \citet{ramanujan_deformation_1998} and \citet{sui_hybrid_2008} have used a similar approach, starting from a regular octahedron. The subdivision scheme starts at creating a new node at the middle of each edge. Those initially 30 new nodes (in case of an icosahedron) are then radially shifted until they are located on the circumsphere of the body and connected to form additional faces. This procedure can be repeated numerous times. It is illustrated in Fig.\ \ref{fig:subdivision}. It has to be noted that each subdivision step increases the number of faces according to $N_f^m = N_f^0 \cdot 4^m$, where $m$ is the number of subdivisions and $N_f^0 = 20$ for an icosahedron and $8$ for an octahedron. Although the resulting mesh has surpassing properties in terms of edge length, face area, and angle distributions, one cannot create a mesh with an arbitrary number of faces. This restriction can somewhat be relaxed by starting from another body of lower symmetry and another number of faces.

In Tab.\ \ref{tab:mesh_quality}, the properties of the meshes created by the presented methods are listed and compared for the case of a sphere with $N_f \approx 1280$ faces. The meshes are illustrated in Fig.\ \ref{fig:meshtypes}. Obviously, the subdivided mesh has the smallest scatter in face area, edge length, normal-to-normal angle, and edge-to-edge angle distributions.  By default, we use the mesh obtained from the icosahedron if not elsewise stated.


\section{Theory}
\label{sec:theory}

\begin{figure}
\centering
\begin{tikzpicture}
\draw[rotate=40,fill=lightgray,thick] (0,0) ellipse (2cm and 1.5cm);
\draw[-latex,very thick] (-3,0)--coordinate[pos=0.6](anker)(3,0) node[pos=0.6,above right]{$\theta$} node[right]{$x$};
\draw[rotate=40,thick] (-2,0)--(2,0) node[pos=0.85,above left]{$r_a$};
\draw[rotate=40,thick] (0,-1.5)--(0,1.5) node[pos=0.8, above right]{$r_c$};
\draw[thick] (anker) arc (0:40:0.6cm);
\draw[-latex,line width=3pt,darkgray] (-1,2.2)--(2,2.2) node[below right]{$\dot \gamma$};
\draw[-latex,line width=3pt,darkgray] (1,-2.2)--(-2,-2.2);
\draw[-latex,very thick,rotate=40] (-0.7,1.8) to[out=12,in=168] (0.7,1.8) node[midway,above left]{$\omega$};
\end{tikzpicture}
\caption{Sketch of the tank-treading geometry. The capsule cross-section is shown in the $xy$-plane. It is deformed with major and minor semiaxes $r_a$ and $r_c$. The inclination angle $\theta$ is taken between the major semiaxis and the $x$-axis (velocity direction of the external flow). The membrane rotates about its spatially fixed shape with angular velocity $\omega$. The flow direction of the unperturbed ambient fluid is also shown (dark gray arrows).}
\label{fig:geometry}
\end{figure}
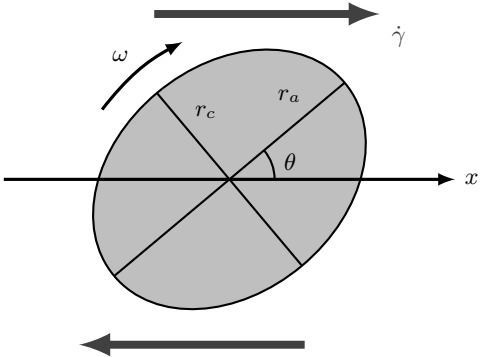

We assume that the ambient fluid and the fluid inside the capsule are Newtonian and have the same properties, especially the same density ($\rho = \rho_{\text{in}} = \rho_{\text{out}}$) and viscosity ($\lambda = \eta_{\text{in}} / \eta_{\text{out}} = 1$ and $\eta = \eta_{\text{in}} = \eta_{\text{out}}$). This commonly adopted assumption \cite{kraus_fluid_1996, navot_elastic_1998, sui_hybrid_2008} significantly simplifies the computations without losing too much generality. In the limit of Stokes flow, the capsule Reynolds number
\be
\text{Re} = \f{\rho \dot \gamma r^2}{\eta}
\ee
is small, and inertia effects can be neglected. The only physical parameter left is the dimensionless shear rate
\be
\label{eq:shearrate}
G = \f{\dot \gamma \eta r}{k_s},
\ee
where $\dot \gamma$ is the shear rate of the unperturbed ambient fluid, and $r$ and $k_s$ are the radius and the surface elastic shear modulus of the initially spherical capsule.

Due to the presence of the external shear flow, the membrane deforms. The generated membrane tensions oppose the shear forces exerted by the fluid. For $G \ll 1$, a small membrane deformation suffices to compensate the shear forces, and the capsule shape is only slightly perturbed. In a simple shear flow, after an initial transient, steady tank-treading motion of an initially spherical capsule, deformed to an ellipsoid, is observed \cite{barths-biesel_motion_1980, barths-biesel_time-dependent_1981, keller_motion_1982, ramanujan_deformation_1998, kraus_fluid_1996}. Tank-treading has been described first by \citet{schmid-schonbein_fluid_1969}. The shape and dynamics of the capsule can then be defined by three constant parameters: the Taylor deformation parameter $D$, the inclination angle $\theta$, and an angular velocity $\omega$. The stationary geometry is shown in Fig.\ \ref{fig:geometry}.

The inclination angle $\theta$ of the membrane is taken between its largest semiaxis $r_a$ and the $x$-axis, the direction of the fluid velocity. \citet{dupin_modelingflow_2007} have extracted $\theta$ from an ellipsoid fit of the membrane cross-section on the $xz$-plane. A common alternative is the comparison of the capsule shape with an ellipsoid with the same inertia tensor $\bm I$. Following \citet{ramanujan_deformation_1998}, its components are given by
\be
I_{\alpha \beta} = \f 15 \sum_i^{\text{faces}} A_i \left(\bm r_i^2 \delta_{\alpha \beta} r_\gamma - r_{i\alpha} r_{i\beta} r_{i\gamma}\right) n_{i\gamma}
\ee
where $\bm r_i$ is the centroid of face $i$ and $\bm n_i$ its normal. The diagonalized inertia tensor of an ellipsoid with unit mass and constant density is $\bm I = \text{diag} (r_b^2 + r_c^2, r_a^2 + r_c^2, r_a^2 + r_b^2) / 5$. $r_a$ and $r_c$ are the largest and smallest semiaxis of the ellipsoid, and $r_b$ is the intermediate semiaxis. Assuming that the symmetry axes of the deformed capsule coincide with the principal axes of the inertia tensor, the ellipsoid's principal semiaxes and the inclination angle can be computed. For small deformations, this is an excellent approximation.

The Taylor deformation index is defined as
\be
D = \f{r_a - r_c}{r_a + r_c} > 0.
\ee
It is a measure for the deformation of the capsule, and $D = 0$ holds for a sphere. In the case of small deformations in Stokes flow ($G, \text{Re} \ll 1$), the shape of the capsule can be described analytically \cite{barths-biesel_time-dependent_1981}. In simple shear flow in the steady state, the relation between the dimensionless shear rate $G$ and the deformation parameter $D$ is
\be
\label{eq:expectationvalue}
D = \f 54 \f{3 \alpha_2 + 4 \alpha_3}{\alpha_1 (3 \alpha_2 + 5 \alpha_3) + 2 \alpha_3 (\alpha_2 + \alpha_3)} G + O(G^2).
\ee
The coefficients $\alpha_1$, $\alpha_2$, and $\alpha_3$ can be extracted from the constitutive model, Eq.\ (\ref{eq:constitutivelaw}), by expansion,
\begin{widetext}
\be
\label{eq:energyexpansion}
w^S / k_s = w_0 + \alpha_1 \Lambda_a + \f 12 (\alpha_2 + \alpha_1) \Lambda_a^2 + \alpha_3 (\Lambda_b - \Lambda_a) + O(\Lambda_a^3, \Lambda_a \Lambda_b, \Lambda_b^2),
\ee
\end{widetext}
where $\Lambda_a = \ln(\lambda_1 \lambda_2)$ and $2 \Lambda_b = \lambda_1^2 + \lambda_2^2 - 1$. This leads to $\alpha_1 = 0$, $\alpha_2 = 2/3$, and $\alpha_3 = 1/3$ for the neo-Hookean law and for Skalak's law, Eq.\ (\ref{eq:constitutivelaw}), with $k_s = k_\alpha$. \citet{barths-biesel_motion_1980} has also found an analytic expression for the deviation of the inclination angle at small $G$,
\be
\label{eq:analyticangle}
\f{\theta^o}{\pi} = \f{\pi / 4 - \theta}{\pi} = \f{15}{8} G
\ee
for a neo-Hookean membrane and Skalak's law with $k_s = k_\alpha$. In the limit $G \to 0$, the angle approaches $\theta = \pi / 4$. The opposite angle $\theta^o$ is the relevant angle in this problem since it is proportional to $G$. It is the deviation from the inclination angle of a stiff sphere, where $\theta = \pi / 4$.

In the steady state, the membrane nodes rotate about the fixed shape of the capsule. This tank-treading behavior can be quantified by the rotation period $T$ or the angular velocity $\omega$ in the steady state. \citet{kraus_fluid_1996} have measured the time between two successive vertex crossings of the $xy$-plane. This time corresponds to half the rotation period. However, this value is time integrated and cannot resolve possible numerical fluctuations of the angular velocity. Another approach is the direct computation of the angular velocity of a membrane node by $\omega_i = \Delta \varphi_i / \Delta t$, where $\Delta \varphi_i$ is the angle swept by that node projected on the $xz$-plane during time $\Delta t$. For a stiff sphere, the angular velocity is $\omega = \dot \gamma / 2$, but $\omega < \dot \gamma / 2$ holds for deformable capsules.

Additionally to the approach using the inertia tensor, we have computed the deformed capsule shape by a linear fit of the membrane node positions. Minimizing
\be
\label{eq:chisquare}
\chi^2 = \sum_i^{\text{nodes}} \left(\zeta_x x_i^2 + \zeta_y y_i^2 + \zeta_z z_i^2 - 2 \zeta_{xz} x_i z_i - 1\right)^2
\ee
with respect to the fit parameters $\zeta_x$, $\zeta_y$, $\zeta_z$, and $\zeta_{xz}$ and diagonalizing the matrix
\be
Q = \begin{pmatrix} \zeta_x & 0 & -\zeta_{xz} \\ 0 & \zeta_y & 0 \\ -\zeta_{xz} & 0 & \zeta_z \end{pmatrix},
\ee
one can directly compute the semiaxes, the deformation parameter, and the inclination angle about the $y$-axis. For convenience, we have not allowed rotations about the $x$- or the $z$-axis in Eq.\ (\ref{eq:chisquare}). Comparing the results obtained from the inertia tensor and the linear fit, we observe that both the inclination angle and the deformation parameter are virtually identical. For that reason, we will show the data obtained from the inertia tensor only. However, the values of the semiaxes $r_a$, $r_b$, and $r_c$ are slightly underestimated by the inertia tensor. The reason is the discretization of the mesh and the use of flat triangular elements. Since the semiaxes are not explicitly required for the characterization of the deformation and the values of $D$ and $\theta$ are correct, we have not attempted to improve the results obtained by the inertia tensor method.


\section{Simulations and results}
\label{sec:simulations}

All distances are made dimensionless using the lattice constant $\Delta x$ as characteristic length scale. As mentioned above, from the constitutive law, Eqs.\ (\ref{eq:constitutivelaw}) and (\ref{eq:energyexpansion}), the expansion parameters $\alpha_1 = 0$, $\alpha_2 = 2/3$, and $\alpha_3 = 1/3$ follow for $k_s = k_\alpha$. Inserting these values in Eq.\ (\ref{eq:expectationvalue}), one obtains $D/G = 25/4$ in the linear regime. We restrict ourselves to the Skalak model, Eq.\ (\ref{eq:constitutivelaw}), with $k_s = k_\alpha$ in all the simulations. If not otherwise stated, the lattice Boltzmann relaxation parameter is set to $\tau = 1$, and thus the time step scales like $\Delta t \propto \Delta x^2$ (diffusive scaling). The simulation box is a cube with $H^3$ fluid lattice nodes. The capsule with initial radius $r$ is placed at the center of the box. We choose the $x$-axis as the velocity direction and the $y$-axis as the vorticity direction. The velocity gradient is along the $z$-axis. The bounding plates are located at $z = \pm H/2$.

In Sec.\ \ref{subsec:finite_size}, we first determine the effects of the relative simulation box size $H/r$, the Reynolds number $\text{Re}$, the reduced shear rate $G$, and the LBM relaxation parameter $\tau$ on the deformation state of the capsule. Based on those first results, the influence of the membrane tessellation is analyzed in Sec.\ \ref{subsec:meshdiscretization}. In Sec.\ \ref{subsec:interpolation}, we test the influence of the interpolation stencils on the numerical results and specify the convergence behavior of the IBLBFEM scheme.

\subsection{General simulation parameters}
\label{subsec:finite_size}

\begin{figure*}
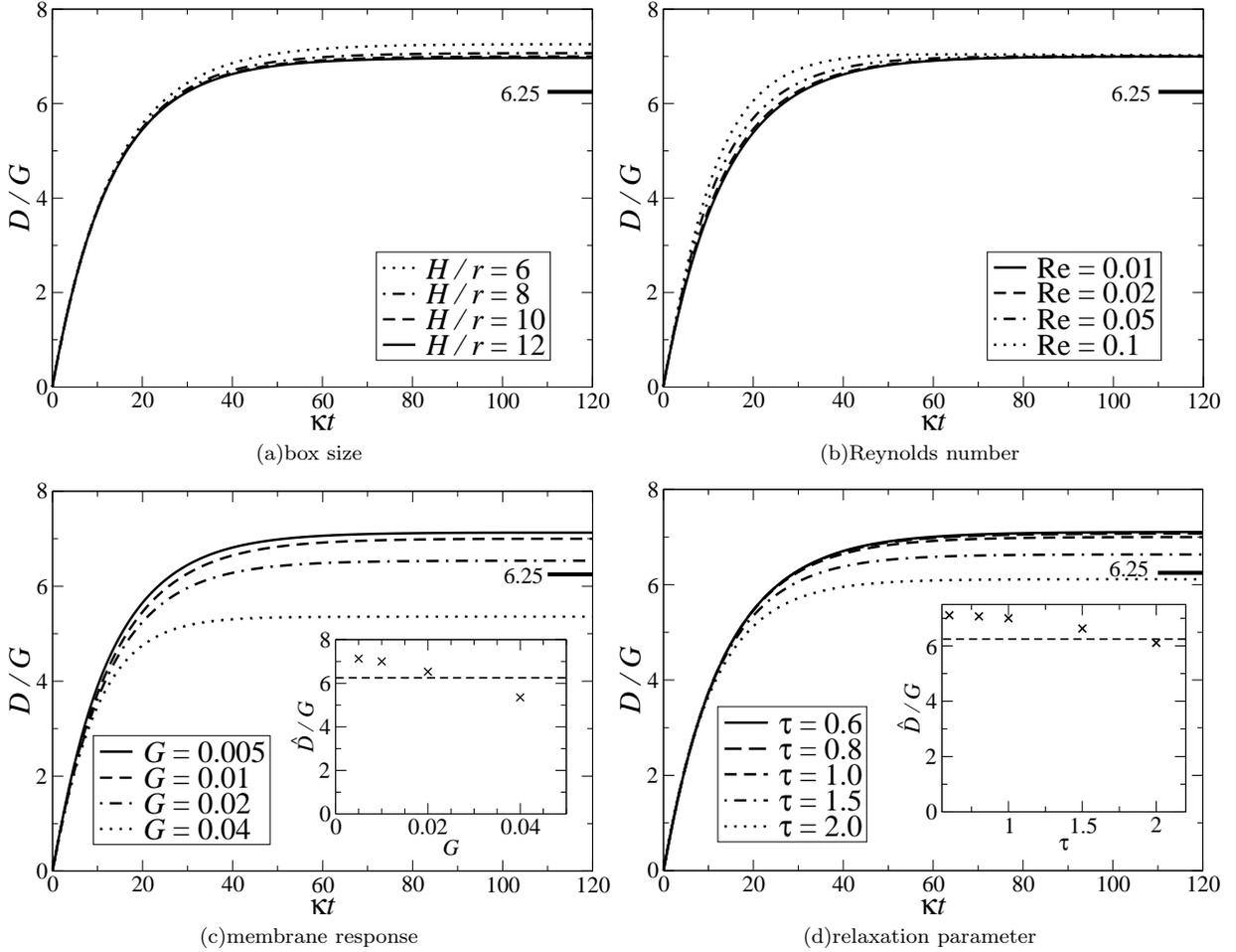

\centering
\subfigure[\label{fig:finitesize}box size]{\includegraphics[width=0.45\linewidth,clip=true]{finite_size}}
\subfigure[\label{fig:inertia}Reynolds number]{\includegraphics[width=0.45\linewidth,clip=true]{inertia}}
\subfigure[\label{fig:linearmembrane}membrane response]{\includegraphics[width=0.45\linewidth,clip=true]{nonlinearity}}
\subfigure[\label{fig:tau}relaxation parameter]{\includegraphics[width=0.45\linewidth,clip=true]{tau}}
\caption{Influence of finite size, inertia, nonlinear membrane response, and LBM relaxation parameter on the deformation parameter, which is shown as a function of the reduced time $\kappa t$, where $\kappa = \dot \gamma / G$. \subref{fig:finitesize} finite size: $\text{Re} = 0.02$, $G = 0.01$, $\tau = 1$, and $H/r = 6, 8, 10, 12$; \subref{fig:inertia} inertia: $H/r = 10$, $G = 0.01$, $\tau = 1$, and $\text{Re} = 0.01, 0.02, 0.05, 0.1$; \subref{fig:linearmembrane} nonlinear membrane response: $\text{Re} = 0.02$, $H/r = 10$, $\tau = 1$, and $G = 0.005, 0.01, 0.02, 0.04$; \subref{fig:tau} relaxation parameter: $\text{Re} = 0.02$, $H/r = 10$, $G = 0.01$, and $\tau = 0.6, 0.8, 1.0, 1.5, 2.0$. In all simulations, $N_f = 1280$, $r = 5$, and $\phi_4$ have been used. The influence of the finite box is negligible for $H/r \geq 10$, and inertia effects are unimportant for $\text{Re} \leq 0.02$. The inset in subfigure \subref{fig:linearmembrane} shows the final value of $D/G$ versus the reduced shear rate $G$. The dashed line is the analytic result for small deformations. For $G \geq 0.02$, nonlinear effects become obvious, but $G = 0.01$ is a good approximation for a small deformation. The relaxation parameter $\tau$ should not be much larger than unity. It can be seen from the inset of subfigure \subref{fig:tau} that the final value of $D/G$ strongly depends on $\tau$, if $\tau > 1$. The theoretical value, $D / G = 6.25$, is shown in all subfigures. One notices that the deformation parameters consequently are too large.}
\label{fig:boxsize}
\end{figure*}

We note that, in the limit $\text{Re} \ll 1$, $H/r \gg 1$, and $G \ll 1$, all curves $D(\kappa t)/G$, with $\kappa = \dot \gamma / G$, collapse on the same master curve. This can be used to study the impact of finite size effects ($H/r \not \gg 1$), inertia ($\text{Re} \not \ll 1$), nonlinear contributions ($G \not \ll 1$) from the constitutive membrane model, and the LBM relaxation parameter $\tau$ by inspecting the deviations from the master curve. For all simulations in this section, the 4-point stencil $\phi_4$, Eq.\ (\ref{eq:4point}), and an icosahedron-based mesh with $N_f = 1280$ faces and $r = 5$ have been used.

Similar to Sui et al.\ \cite{sui_hybrid_2008}, we have first determined the minimum system size in terms of $H/r$ to safely neglect self-interaction of the capsule or interactions with the walls. The simulation parameters are $\text{Re} = 0.02$, $G = 0.01$, and $\tau = 1$. We have examined the cases $H/r = 6, 8, 10$, and $12$. The resulting time evolution of the Taylor parameter is shown in Fig.\ \ref{fig:finitesize}. We come to the same conclusion as Sui et al.\ \cite{sui_hybrid_2008} that a box size of $H/r = 10$ is sufficient for modeling unbounded simple shear flow. The difference between the plateau values of the Taylor parameter for $H/r = 10$ and $12$ is less than $0.5\%$. For this reason, the system size is taken to be $H/r = 10$ in all following simulations. However, it is obvious that the deformation parameters are too large. The expected value, $D / G = 6.25$, is shown in Fig.\ \ref{fig:boxsize}. We will see in this section that this is not related to effects caused by inertia, nonlinear membrane response, or the relaxation parameter. This observation will be discussed in more detail in the following sections, and the explanation will be given in Sec.\ \ref{subsec:interpolation}.

In the next step, we have investigated the validity of the Stokes flow assumption. A series of simulations with $H/r = 10$, $G = 0.01$, and $\tau = 1$ has been carried out. The tested Reynolds numbers are $\text{Re} = 0.01$, $0.02$, $0.05$, and $0.1$. Sui et al.\ \cite{sui_hybrid_2008} have observed that inertia effects are small up to a Reynolds number of about $0.025$. We come to a similar conclusion. In Fig.\ \ref{fig:inertia}, the deformation parameter is shown for different $\text{Re}$. Differences between $\text{Re} = 0.01$ and $0.02$ can only be noticed slightly in the transient evolution. Since we are interested in the steady-state behavior, the Reynolds number will be set to $0.02$ in all subsequent simulations. We have also studied the time evolution if the equilibrium in Eq.\ (\ref{eq:equilibrium}) is linearized, i.e., substituted by
\be
f_i^{\text{eq}} = w_i\, \rho \left(1 + 3 \bm c_i \cdot \bm u\right).
\ee
One can show that this leads to the removal of the advection term $\bm u \cdot \nabla \bm u$ in the Navier-Stokes equations. Still, the partial time derivative $\partial \bm u / \partial t$ is present. Within the kinematic framework of LBM, the time derivative cannot be removed, and exact Stokes flow cannot be simulated. We have observed that the time evolution of the deformation parameter does not change when the advection term is removed with respect to the cases where the full equilibrium has been considered (data not shown). We thus assume that the Reynolds number effects visible in Fig.\ \ref{fig:inertia} are due to the partial time derivative and not the advective term. Still, neglecting the second-order term in the equilibrium, the computing time for LBM could be decreased significantly ($\approx 25\%$). This observation can be useful for high performance simulations at small Reynolds numbers. However, since we have tested the first-order equilibrium afterwards, for the remaining simulations the second-order equilibrium has been employed.

Additionally, we have tested up to which value of the reduced shear rate $G$ the linearity assumption, Eq.\ (\ref{eq:expectationvalue}), is valid. In Fig.\ \ref{fig:linearmembrane}, the time evolution of $D$ is shown. The simulation parameters are $\text{Re} = 0.02$, $H/r = 10$, and $\tau = 1$. The reduced shear rates are $G = 0.005$, $0.01$, $0.02$, and $0.04$. We find that $G = 0.01$ is sufficient to ensure the validity of the linear approximation. For $G \geq 0.02$, the deviations become significant, and second-order correction terms should be included. In all the following simulation, we have chosen $G = 0.01$. Note that the plateau values of $D/G$ between $G = 0.005$ and $0.01$ differ by less than $2\%$.

It is known that the BGK LBM relaxation parameter $\tau$ plays a critical role in the correct setup of the simulations. Both the accuracy of the bulk LBM and the no-slip bounce-back boundary conditions depend on its value \cite{he_analytic_1997, holdych_truncation_2004, kruger_shear_2009}. Recently, \citet{le_boundary_2009} reported that combined IBM-LBM simulations are strongly affected by the magnitude of $\tau$, especially if $\tau > 1$. We address this issue by comparing simulations with different values of the relaxation parameter for $\text{Re} = 0.02$, $H/r = 10$, and $G = 0.01$. The results are shown in Fig.\ \ref{fig:tau}. It can clearly be seen that the results are relatively independent of $\tau$ for $\tau \leq 1$. If $\tau$ becomes significantly larger than unity, the solutions start to diverge. This effect may be related to the single relaxation time (BGK) LBM which has been employed. Having the efficiency of the simulations in mind, it is desired to keep the simulation time as short as possible. This can be achieved by increasing the time step $\Delta t$. From
\be
\label{eq:timestep}
\nu = \f{\tau - 1/2}{3} \f{\Delta x^2}{\Delta t}
\ee
we can see that increasing $\tau$ also increases $\Delta t$, if the kinematic viscosity $\nu$ and the spatial resolution $\Delta x$ are kept fixed. This reduces the number of necessary time steps in the simulations. However, if $\tau$ becomes too large, drastic numerical artifacts appear, and the simulations become unreliable. It seems to be a compromise to choose $\tau \approx 1$.

Summing up, we come to the first conclusion that a cubic box of length $H/r = 10$, a Reynolds number of $\text{Re} = 0.02$, a reduced shear rate of $G = 0.01$, and a LBM relaxation parameter of $\tau = 1$ are excellent approximations to the unbound Couette flow at vanishing Reynolds number in the linear elastic limit. However, the deformation parameters are larger than expected. Theory predicts $D/G = 25/4 = 6.25$ for the Skalak membrane with $k_s = k_\alpha$, but typical values are $D/G \approx 7.0$ in the presented simulations. The expectation value is also shown in Fig.\ \ref{fig:boxsize}. There are two possible reasons for this discrepancy: first, the membrane mesh may be too coarse. Second, the interpolation and spreading, required for the coupling between Lagrangian and Eulerian meshes, may introduce numerical artifacts which effectively lead to a softer or, equivalently, larger capsule. We stress that the IBM generates an artificial length scale $L_I$ related to the width of the interpolation. If this length scale is not small with respect to the length $L_M$ of significant changes in fluid velocity and membrane tensions, a detrimental effect of the interpolation and spreading is expected. In Sec.\ \ref{subsec:meshdiscretization}, we first test the effect of the mesh discretization. Afterwards, in Secs.\ \ref{subsec:interpolation}, we turn to the influence of the IBM interpolation stencil on the simulations.

\subsection{Mesh discretization}
\label{subsec:meshdiscretization}

\begin{figure*}
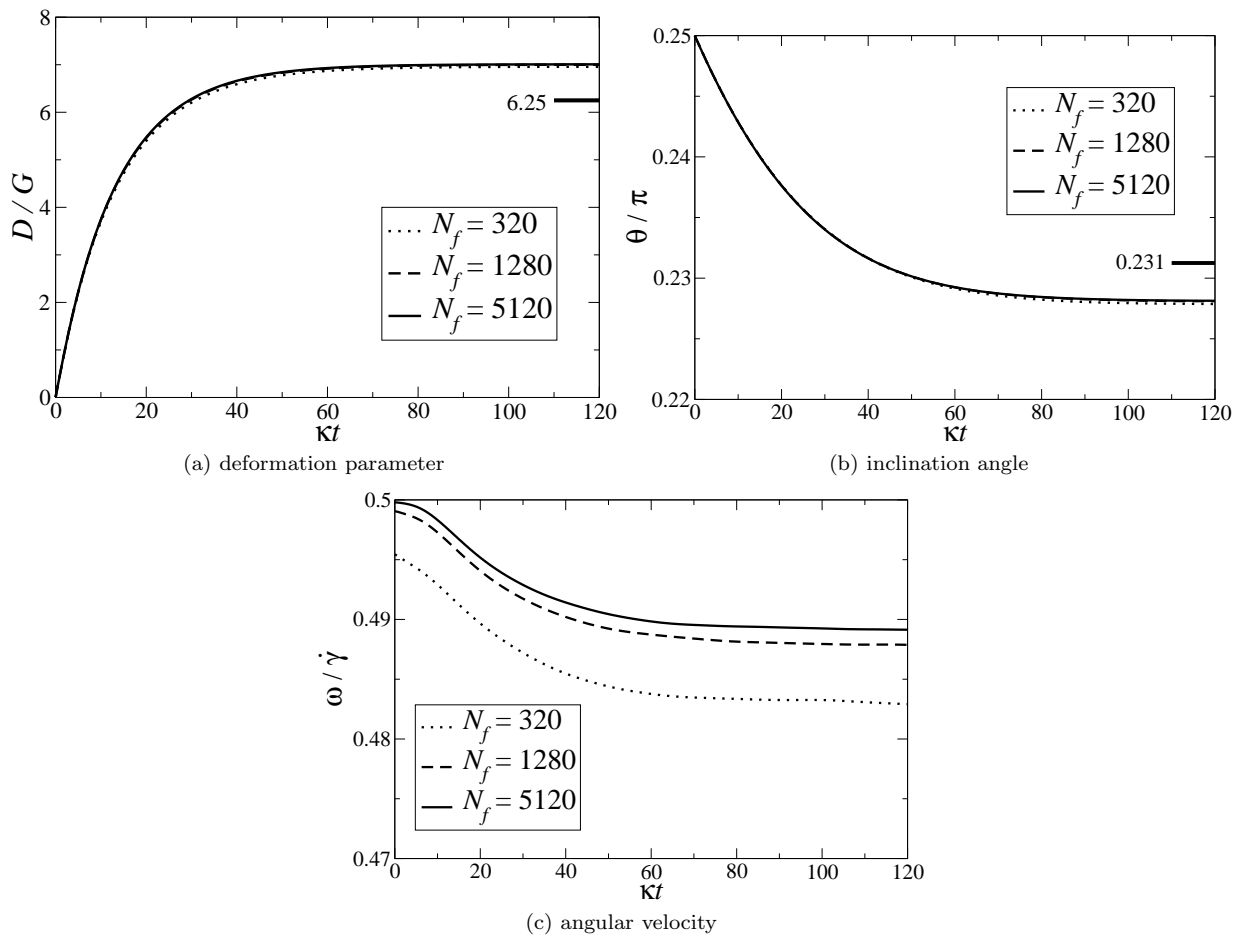

\centering
\subfigure[\label{fig:defparameter} deformation parameter]{\includegraphics[width=0.45\linewidth,clip=true]{mesh_res_D}}
\subfigure[\label{fig:inclangle} inclination angle]{\includegraphics[width=0.45\linewidth,clip=true]{mesh_res_theta}}
\subfigure[\label{fig:angvel} angular velocity]{\includegraphics[width=0.45\linewidth,clip=true]{mesh_res_omega}}
\caption{Effect of the icosahedron-based mesh resolution on \subref{fig:defparameter} the deformation parameter, \subref{fig:inclangle} the inclination angle, and \subref{fig:angvel} the angular velocity. The simulation parameters are $\text{Re} = 0.02$, $H / r = 10$, $G = 0.01$, $\tau = 1$, and $r = 5$. The 4-point interpolation stencil has been used. There is virtually no change in the deformation parameter when the $5120$-element mesh is replaced by the $1280$- or the $320$-element mesh. The same is valid for the inclination angle. The angular velocity is extremely sensitive to the number of face elements, but convergent behavior is evident. The expected analytic values for the deformation parameter and the inclination angle are also shown in the subfigures.}
\label{fig:coarseness}
\end{figure*}

\begin{figure*}
\centering
\subfigure[\label{fig:defparameter2} deformation parameter]{\includegraphics[width=0.45\linewidth,clip=true]{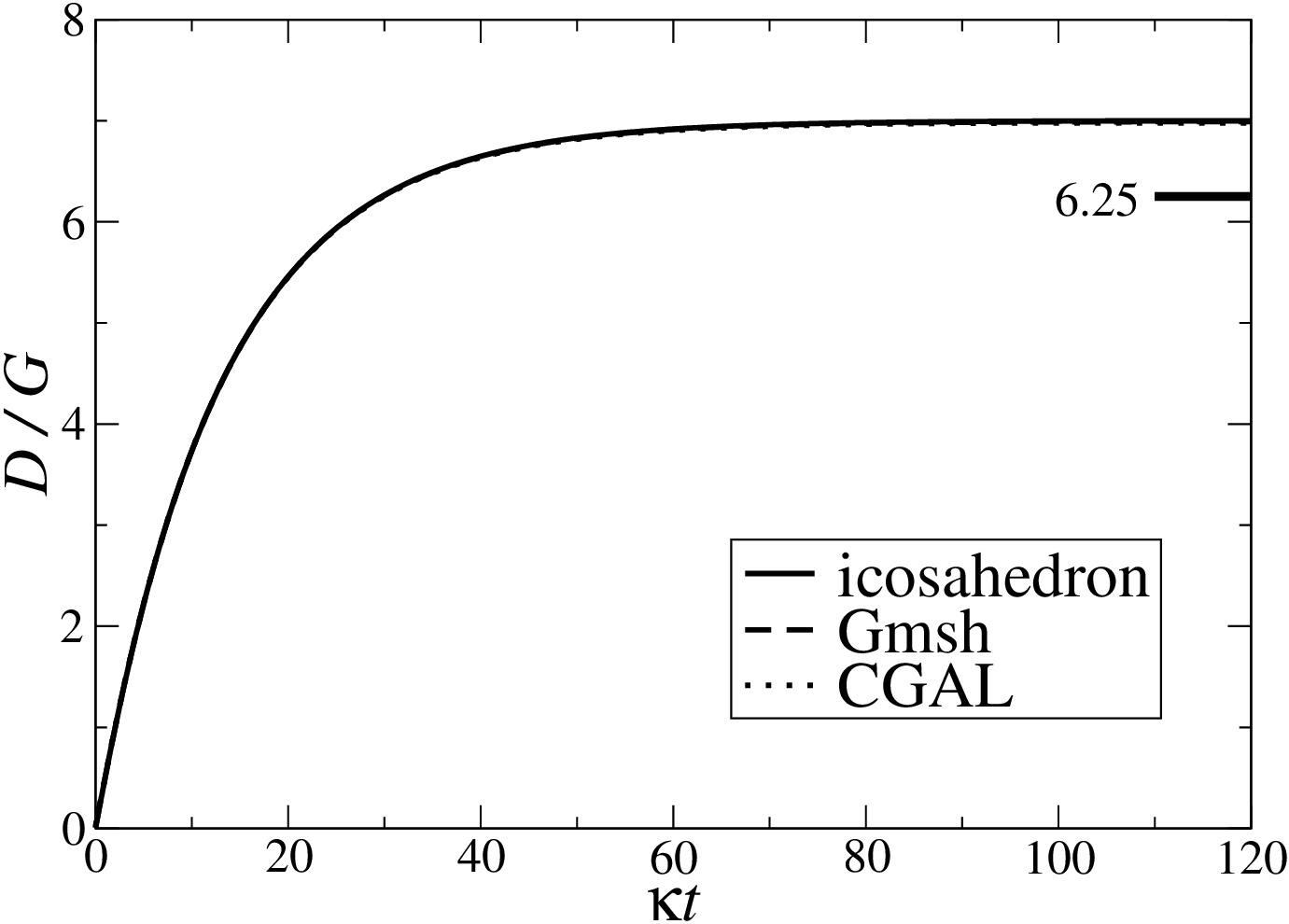}}
\subfigure[\label{fig:inclangle2} inclination angle]{\includegraphics[width=0.45\linewidth,clip=true]{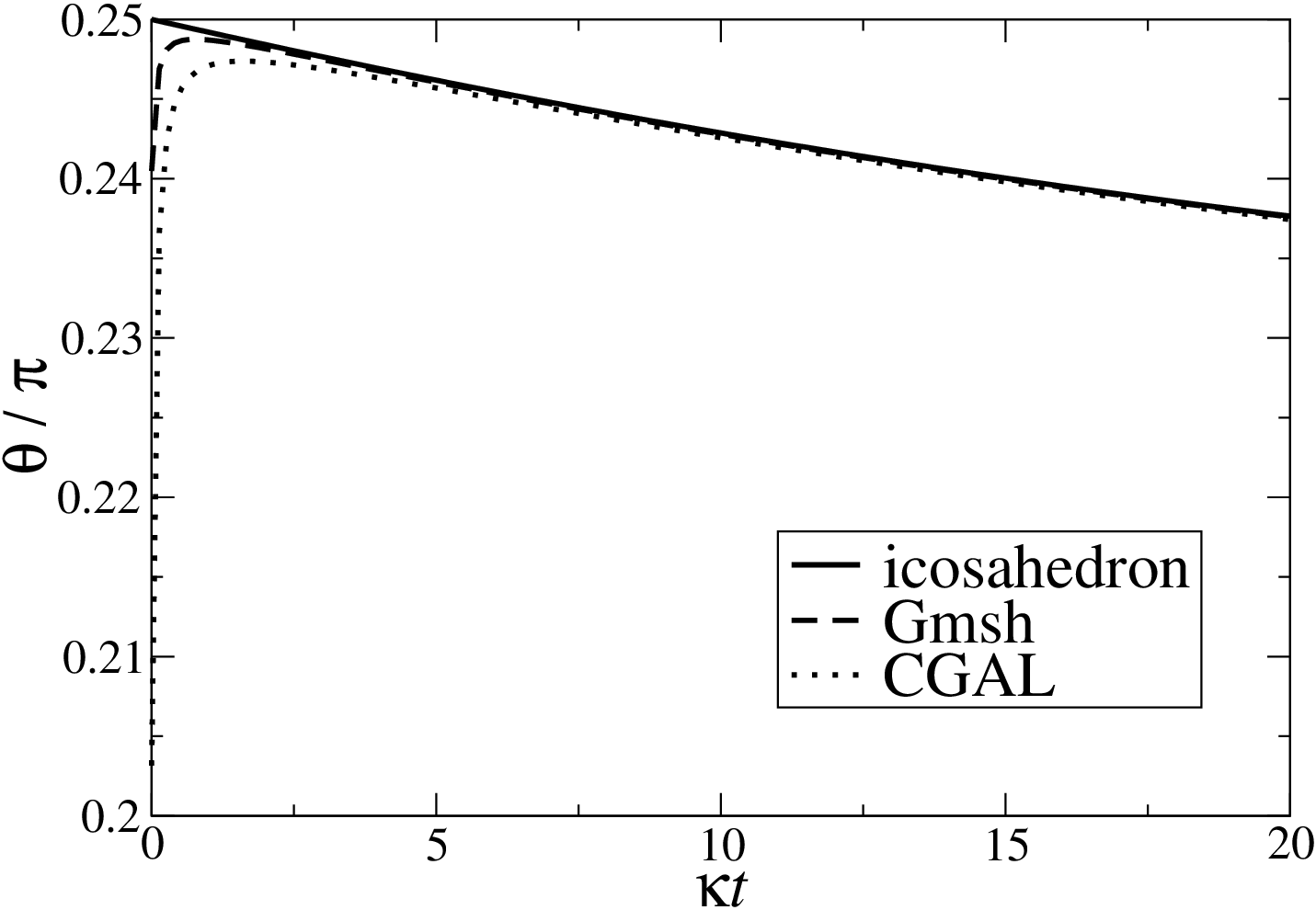}}
\subfigure[\label{fig:angvel2} angular velocity]{\includegraphics[width=0.45\linewidth,clip=true]{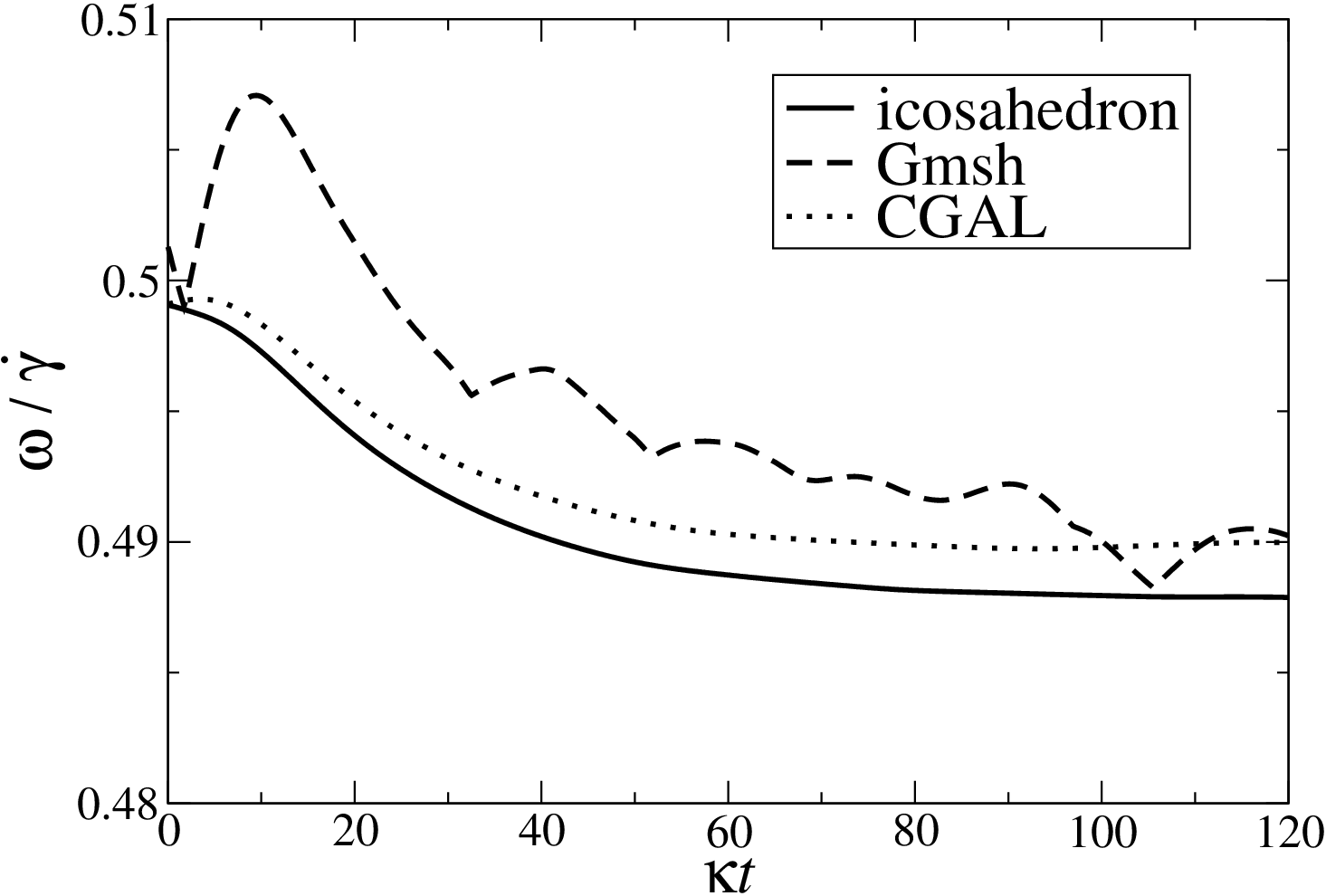}}
\caption{Effect of the tessellation method on \subref{fig:defparameter2} the deformation parameter, \subref{fig:inclangle2} the inclination angle, and \subref{fig:angvel2} the angular velocity. The simulation parameters are the same as those in Fig.\ \ref{fig:coarseness}, but $N_f = 1280$ (icosahedron-based), $1284$ (Gmsh), and $1278$ (CGAL). The deformation parameter is not influenced by the type of the mesh, but the inclination angle and the angular velocity are detrimentally affected. The inclination angle is shown at early times only to reveal the deviations. Due to the reduced homogeneity of the meshes, the correct inclination angle is not captured at small deformations where it should be $\theta / \pi \approx 0.25$. Also the angular velocity shows unphysical behavior, especially for the mesh created with Gmsh.}
\label{fig:meshes}
\end{figure*}

In Sec.\ \ref{subsec:finite_size}, we have stated that the deformation parameters are larger than expected from linear theory. From the discussions there, finite size effects of the simulation box, nonlinear membrane responses, and inertia effects could be excluded. Also the choice of the LBM relaxation parameter cannot be responsible for the deviations. The most probable explanations are spatial discretizations due to the membrane tessellation (cf.\ Sec.\ \ref{subsec:mesh}) or the discrete delta functions for interpolation and spreading, Eqs.\ (\ref{eq:2point})--(\ref{eq:4point}). Since length and time scales in the LBM are strongly coupled \cite{holdych_truncation_2004, kruger_shear_2009}, spatial and temporal discretization errors cannot be studied independently.

In order to test the influence of the details of the mesh resolution, we have performed three simulations with identical parameters ($\text{Re} = 0.02$, $H/r = 10$, $G = 0.01$, $\tau = 1$, $r = 5$, and $\phi_4$), but different meshes (icosahedron-based, $N_f = 320$, $1280$ and $5120$ faces). As a consequence, the average distances between neighboring mesh nodes (i.e., the average edge length of the face elements) differ: $\bar l / \Delta x = 1.50$ for the coarsest, $0.75$ for the intermediate, and $0.38$ for the finest mesh. The results for the deformation parameter, the inclination angle, and the angular velocity are shown in Fig.\ \ref{fig:coarseness}. All curves for $D$ nearly collapse. Thus, the mesh discretization can be dropped as explanation for the numerical softening of the capsules since the difference between the meshes with $N_f = 320$ and $5120$ faces is small. The results for the inclination angles are very similar, but all of them show a deviation from the expected value as well. We will come back to this point in Sec.\ \ref{subsec:interpolation}. A good approximation of the angular velocity $\omega$ can only be provided by a large number of face elements.

At this point, we do not see any reason why the average distance between nodes should be less than $\Delta x / 2$, which is sometimes claimed in the literature \cite{peskin_ibm_2002}. This result is very important from an efficiency point of view. In a simulation of a dense suspension of deformable particles, most of the computing time is required for the IBM interpolation and spreading. Since the number of IBM calculations is proportional to the number of mesh nodes, it is worth examining to which extent the membrane resolution can be reduced at fixed fluid lattice resolution without significantly decreasing the accuracy of the simulations. As a consequence, the computing time could be decreased. It has to be noted that, independent of this result, a finer mesh has to be used when the radius is significantly increased since the average distance between neighboring nodes would become too large eventually.

Additional tests have been performed to analyze the effect of different mesh tessellations (cf.\ Sec.\ \ref{subsec:mesh}). The simulation parameters are as above, but $N_f = 1280$ (icosahedron-based), $1284$ (Gmsh), and $1278$ (CGAL). The results are collected in Fig.\ \ref{fig:meshes}. There is no difference in the evolution of the deformation parameter, but the inclination angle is not correctly captured at small deformations when the Gmsh- and CGAL meshes are used. Strong deviations can be seen in the time evolution of the angular velocity for the mesh created by Gmsh. The reason is that the computation of $\omega$ is most susceptible to numerical artifacts caused by reduced homogeneity of the mesh. However, the general behavior of the capsules is similar, and the deformation state is not strongly influenced by the details of the mesh tessellation.

Concluding this section, it is found that details of the mesh (tessellation method and resolution) do not significantly change the deformation state of the capsule. Thus, the deviation of the deformation parameter from the expected analytic value cannot be caused by the discretization of the capsule membrane. Even a small mesh resolution is sufficient to accurately describe its deformation behavior, at least at small values of $G$, cf.\ Fig.\ \ref{fig:defparameter}. Some details of the capsule are not correctly captured by the Gmsh and CGAL meshes. We will employ the icosahedron-based mesh in all remaining simulations.

\subsection{Interpolation and spreading}
\label{subsec:interpolation}

We have examined the influence of the discrete membrane mesh on the simulations in Sec.\ \ref{subsec:meshdiscretization} where only the 4-point interpolation stencil $\phi_4$ has been employed. The conclusion is that the discrepancy between the observed and predicted steady-state values of the deformation parameter $D$ and inclination angle $\theta$ should be related to the interpolation stencil of the IBM. This open point will now be discussed in more detail. For convenience and clarity, we define a numerical capsule radius $N_r = r / \Delta x$, indicating the number of fluid lattice nodes covered by the actual radius $r$.

A consequence of the presence of the interpolation stencils, Eqs.\ (\ref{eq:2point})--(\ref{eq:4point}), is the finite numerical width of the membrane. Consequently, the computed solutions should converge to the analytic predictions for $L_I / L_M \to 0$. Here, $L_I$ is the length scale associated with the numerical membrane thickness, and $L_M$ can be regarded as the radius of the membrane. The values of $L_I$ for the available interpolation stencils may be identified with $\Delta x$, $3 \Delta x / 2$, and $2 \Delta x$ for $\phi_2$, $\phi_3$, and $\phi_4$, respectively. Since numerical width effects can also be seen in multiphase simulations, we refer to a review about diffuse interface methods by \citet{anderson_diffuse-interface_1998}.

As discussed before, one can find comments on the ratio of mesh and lattice resolution $\bar l / \Delta x$ in the literature. \citet{peskin_ibm_2002} suggests that two adjacent nodes on the membrane mesh should have a mean distance $\bar l \leq \Delta x / 2$. This way, it is argued, no fluid could leak between the `holes' in the mesh. However, a more stringent motivation is missing. In Sec.\ \ref{subsec:meshdiscretization}, we have not found evidence supporting this claim. The results indicate that also an average node distance of $\bar l / \Delta x = 1.5$ may lead to a reasonable accuracy. We stress that a necessity for smaller distances $\bar l$ may arise in the limit of stiff particles or strongly deformed membranes.

Beside the choice of the interpolation stencil $\phi$, the ratio $\bar l / \Delta x$ of the mesh and lattice resolutions is the only freedom left in the IBM. If $\bar l / \Delta x$ becomes too large, fluid will eventually leak through the membrane. On the other hand, if $\bar l$ is chosen to be too small, the spatial hydrodynamic resolution becomes worse since less fluid lattice nodes cover the capsule, and the ratio $L_I / L_M$ increases. Within the framework of IBM, there is no obvious, simple way of estimating the optimum value of $\bar l / \Delta x$. Intuitively, both limiting cases restrict the accuracy of the simulations. Therefore, it is important to understand in which range it is safe to operate.

In order to quantify the dependence of the deformation parameter $D$ and the inclination angle $\theta$ on $L_I / L_M$, we have performed two different kinds of studies, each employing $\phi_2$, $\phi_3$, and $\phi_4$. The simulation parameters always are $\text{Re} = 0.02$, $G = 0.01$, $H/r = 10$, and $\tau = 1$.

\paragraph{Convergence for fixed mesh resolution $N_f$}

\begin{figure*}
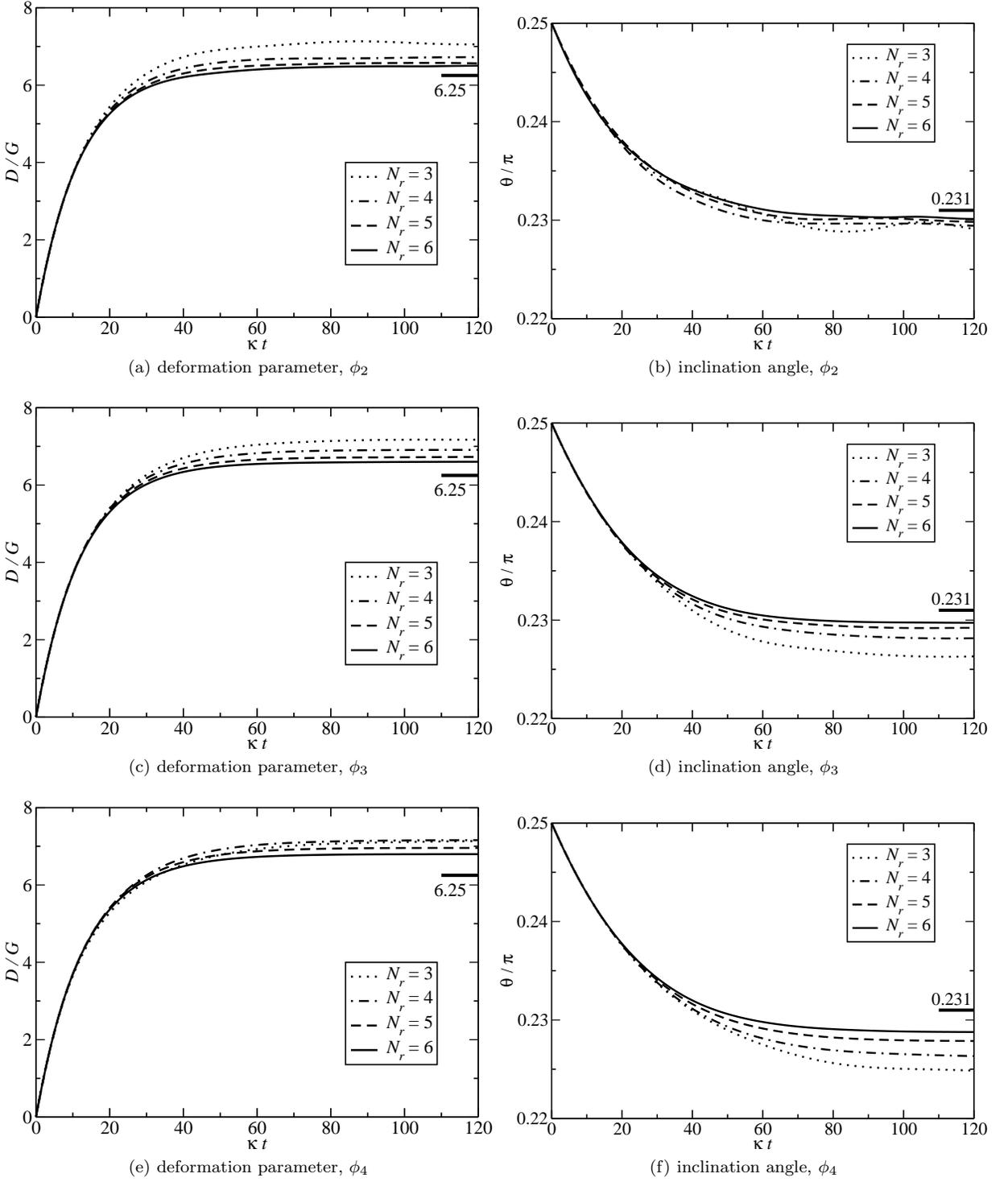

\centering
\subfigure[\label{fig:ratio_320_2_D} deformation parameter, $\phi_2$]{\includegraphics[width=0.45\linewidth,clip=true]{ratio_320_2_D}}
\subfigure[\label{fig:ratio_320_2_theta} inclination angle, $\phi_2$]{\includegraphics[width=0.45\linewidth,clip=true]{ratio_320_2_theta}} \\
\subfigure[\label{fig:ratio_320_3_D} deformation parameter, $\phi_3$]{\includegraphics[width=0.45\linewidth,clip=true]{ratio_320_3_D}}
\subfigure[\label{fig:ratio_320_3_theta} inclination angle, $\phi_3$]{\includegraphics[width=0.45\linewidth,clip=true]{ratio_320_3_theta}} \\
\subfigure[\label{fig:ratio_320_4_D} deformation parameter, $\phi_4$]{\includegraphics[width=0.45\linewidth,clip=true]{ratio_320_4_D}}
\subfigure[\label{fig:ratio_320_4_theta} inclination angle, $\phi_4$]{\includegraphics[width=0.45\linewidth,clip=true]{ratio_320_4_theta}}
\caption{Behavior for varying mesh ratios $\bar l / \Delta x$ for a fixed mesh with $N_f = 320$ faces with the interpolation stencils $\phi_2$ in subfigures \subref{fig:ratio_320_2_D} and \subref{fig:ratio_320_2_theta}, $\phi_3$ in subfigures \subref{fig:ratio_320_3_D} and \subref{fig:ratio_320_3_theta}, and $\phi_4$ in subfigures \subref{fig:ratio_320_4_D} and \subref{fig:ratio_320_4_theta}. $N_r = 3$, $4$, $5$, and $6$ correspond to $\bar l / \Delta x = 0.90$, $1.20$, $1.50$, and $1.80$, respectively.}
\label{fig:ratio_320}
\end{figure*}

\begin{figure*}
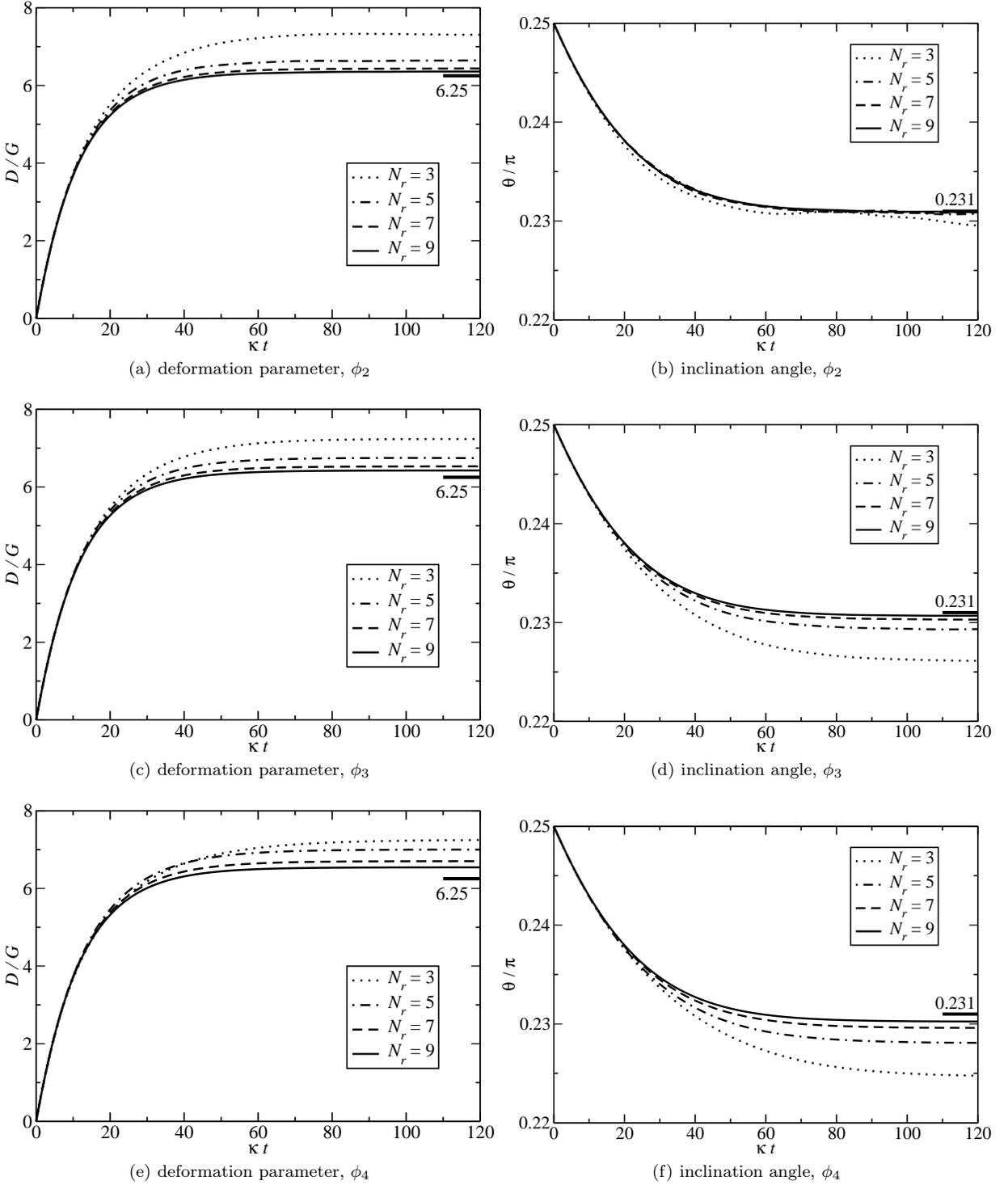

\centering
\subfigure[\label{fig:ratio_1280_2_D} deformation parameter, $\phi_2$]{\includegraphics[width=0.45\linewidth,clip=true]{ratio_1280_2_D}}
\subfigure[\label{fig:ratio_1280_2_theta} inclination angle, $\phi_2$]{\includegraphics[width=0.45\linewidth,clip=true]{ratio_1280_2_theta}} \\
\subfigure[\label{fig:ratio_1280_3_D} deformation parameter, $\phi_3$]{\includegraphics[width=0.45\linewidth,clip=true]{ratio_1280_3_D}}
\subfigure[\label{fig:ratio_1280_3_theta} inclination angle, $\phi_3$]{\includegraphics[width=0.45\linewidth,clip=true]{ratio_1280_3_theta}} \\
\subfigure[\label{fig:ratio_1280_4_D} deformation parameter, $\phi_4$]{\includegraphics[width=0.45\linewidth,clip=true]{ratio_1280_4_D}}
\subfigure[\label{fig:ratio_1280_4_theta} inclination angle, $\phi_4$]{\includegraphics[width=0.45\linewidth,clip=true]{ratio_1280_4_theta}}
\caption{Behavior for varying mesh ratios $\bar l / \Delta x$ for a fixed mesh with $N_f = 1280$ faces with the interpolation stencils $\phi_2$ in subfigures \subref{fig:ratio_1280_2_D} and \subref{fig:ratio_1280_2_theta}, $\phi_3$ in subfigures \subref{fig:ratio_1280_3_D} and \subref{fig:ratio_1280_3_theta}, and $\phi_4$ in subfigures \subref{fig:ratio_1280_4_D} and \subref{fig:ratio_1280_4_theta}. $N_r = 3$, $5$, $7$, and $9$ correspond to $\bar l / \Delta x = 0.45$, $0.75$, $1.06$, and $1.36$, respectively.}
\label{fig:ratio_1280}
\end{figure*}

In this simulation series, we study the influence of the hydrodynamic resolution $N_r$ alone, i.e., we keep the mesh resolution $N_f$ constant, and $\bar l / \Delta x$ changes. This way, it is possible to study the effect of a varying ratio $\bar l / \Delta x$. The employed mesh resolutions are $N_f = 320$ and $1280$.

For the mesh with $320$ faces, we have tested $N_r = 3$, $4$, $5$, and $6$, corresponding to $\bar l / \Delta x = 0.90$, $1.20$, $1.50$, and $1.80$, respectively. The results are shown in Fig.\ \ref{fig:ratio_320}. Although the mesh resolution $N_f$ is small and the mesh ratio $\bar l / \Delta x$ significantly larger than $0.5$, it can be seen that the physics of the system is roughly captured. The accuracy of the solutions increases with a larger magnitude of $N_r$. For $\phi_2$, fluctuations are visible. The smallest radius, $N_r = 3$, yields inaccurate results. In this case, $L_I$ and $L_M$ are comparable. It is interesting to note that even for $\bar l / \Delta x = 1.80$ no detrimental effects appear. The fluid still does not seem to penetrate the membrane.

Additionally, we have tested the mesh with $1280$ faces and $N_r = 3$, $5$, $7$, and $9$, corresponding to $\bar l / \Delta x = 0.45$, $0.75$, $1.06$, and $1.36$, respectively. The results are shown in Fig.\ \ref{fig:ratio_1280}. The overall behavior of the numerical results is similar to those of the series with $N_f = 320$. The smallest radius, $N_r = 3$, is less accurate, and no penetration of the fluid is visible in the presented parameter range.

We have observed that the 2-point interpolation stencil $\phi_2$ fails when $\bar l / \Delta x > 2$ (data not shown). At this point, the spacing between neighboring mesh nodes is so large that fluid can penetrate the capsule membrane. For $\phi_3$ and $\phi_4$, we have not observed a similar behavior at $\bar l / \Delta x = 2$. The probable explanation is the larger range of the interpolations, still keeping the fluid from passing through the membrane.

The above studies strongly suggest that the mesh ratio can be safely chosen somewhere between $0.5$ and $1.5$ without compromising the impermeability of the capsule. This is an important result since it allows us to reduce the computational requirements by a proper choice of $\bar l / \Delta x$ without loss of accuracy.

\paragraph{Convergence for fixed mesh ratio $\bar l / \Delta x$}

\begin{figure*}
\centering
\subfigure[\label{fig:convergence_2_D} deformation parameter, $\phi_2$]{\includegraphics[width=0.45\linewidth,clip=true]{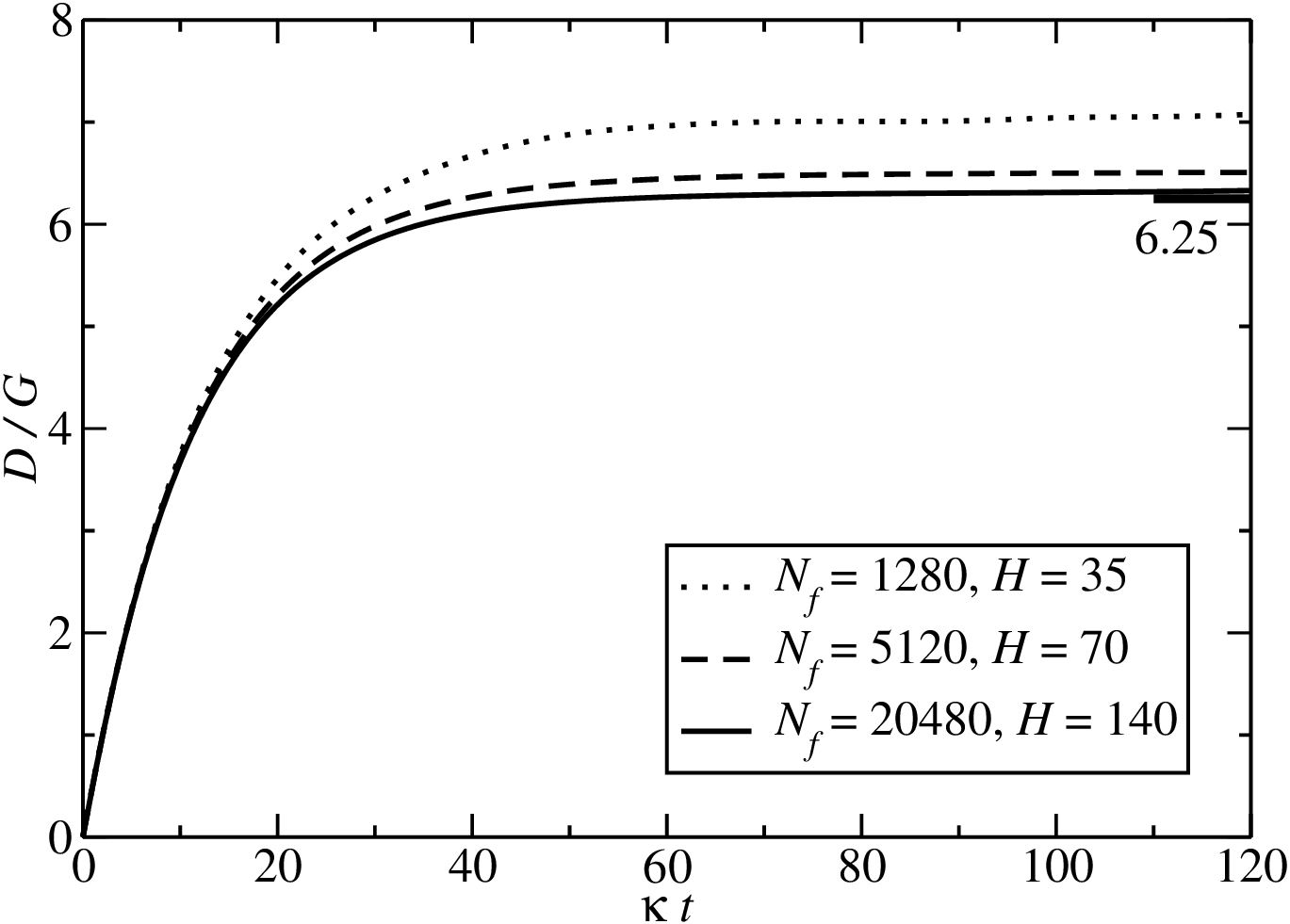}}
\subfigure[\label{fig:convergence_2_theta} inclination angle, $\phi_2$]{\includegraphics[width=0.45\linewidth,clip=true]{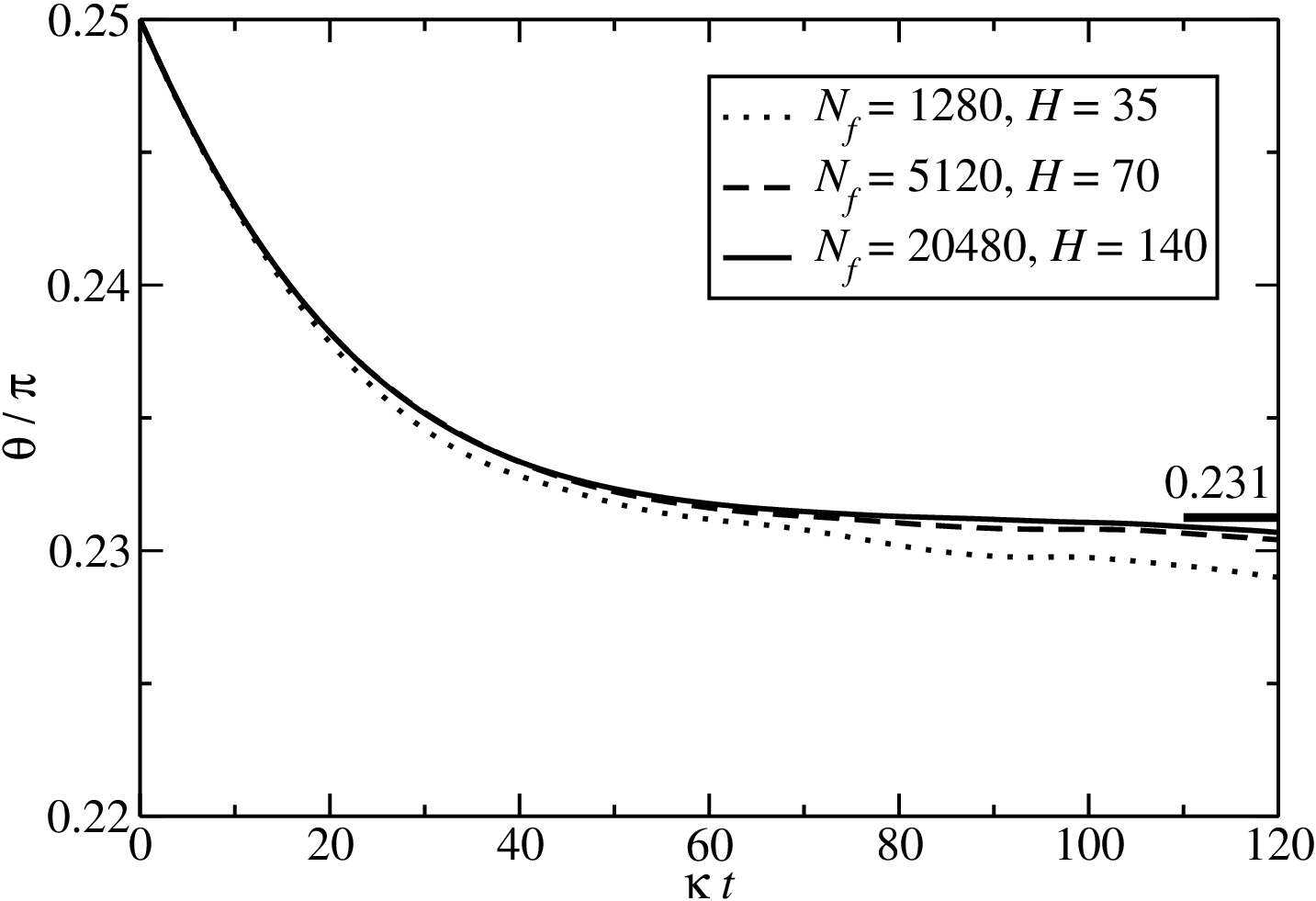}} \\
\subfigure[\label{fig:convergence_3_D} deformation parameter, $\phi_3$]{\includegraphics[width=0.45\linewidth,clip=true]{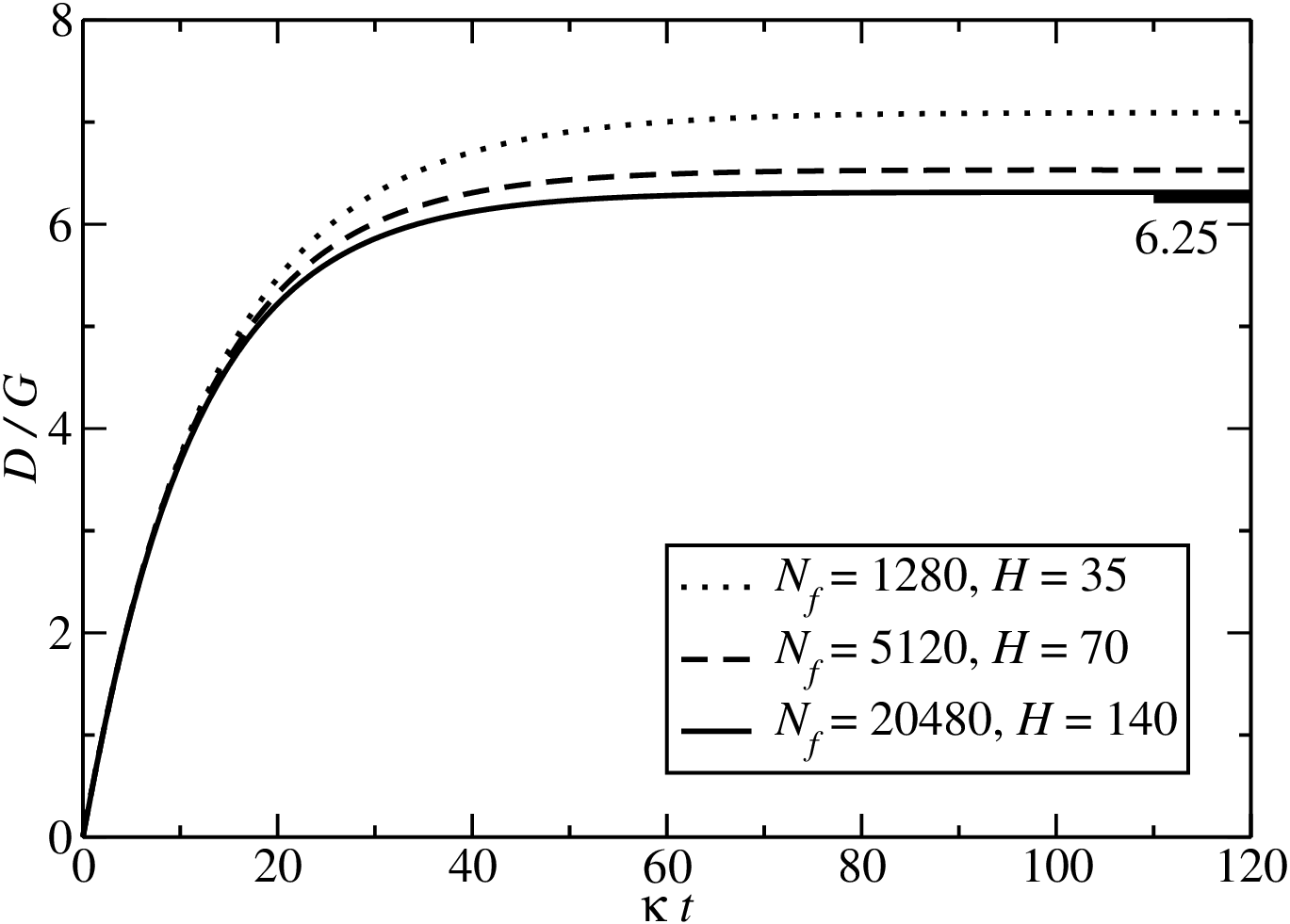}}
\subfigure[\label{fig:convergence_3_theta} inclination angle, $\phi_3$]{\includegraphics[width=0.45\linewidth,clip=true]{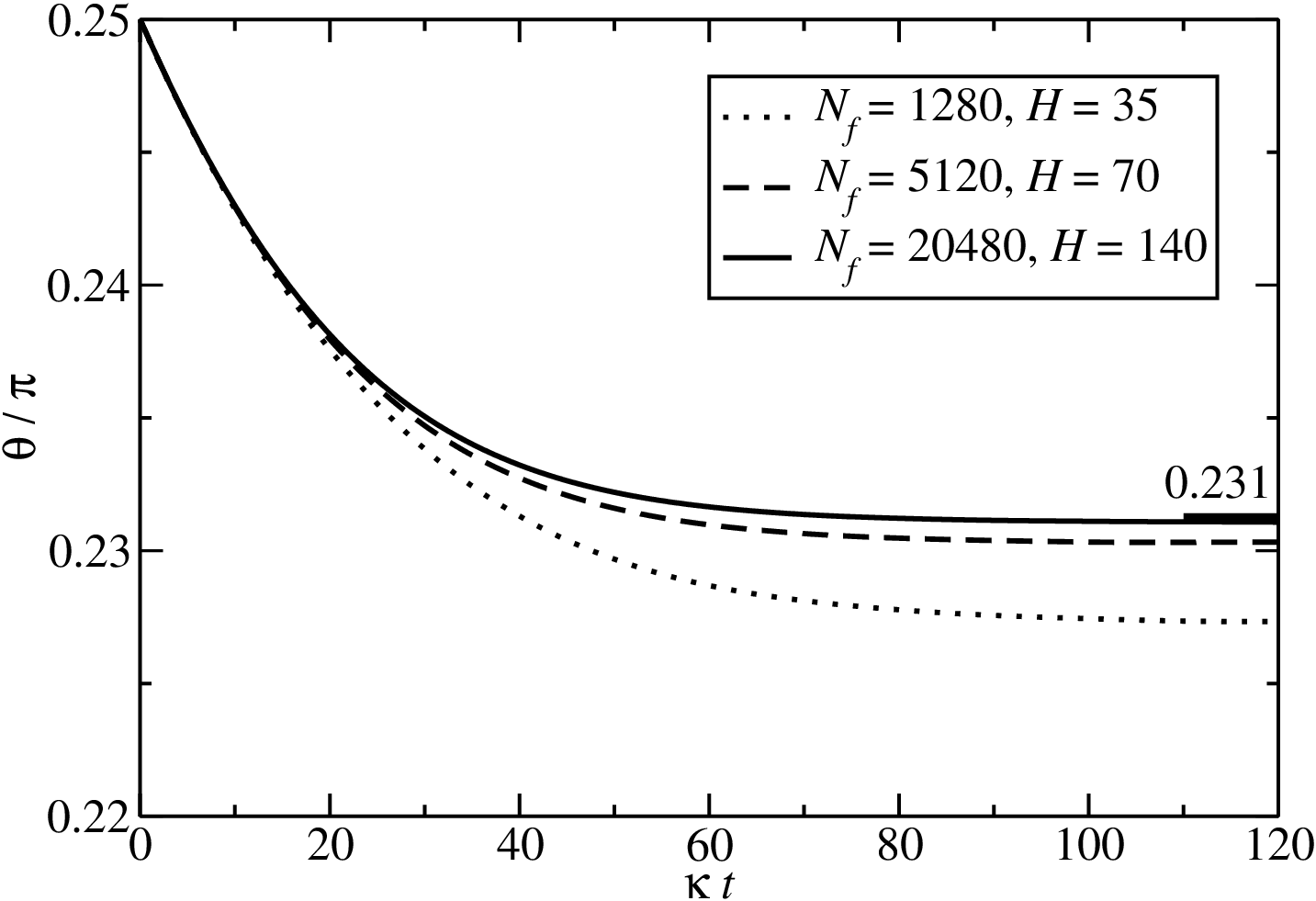}} \\
\subfigure[\label{fig:convergence_4_D} deformation parameter, $\phi_4$]{\includegraphics[width=0.45\linewidth,clip=true]{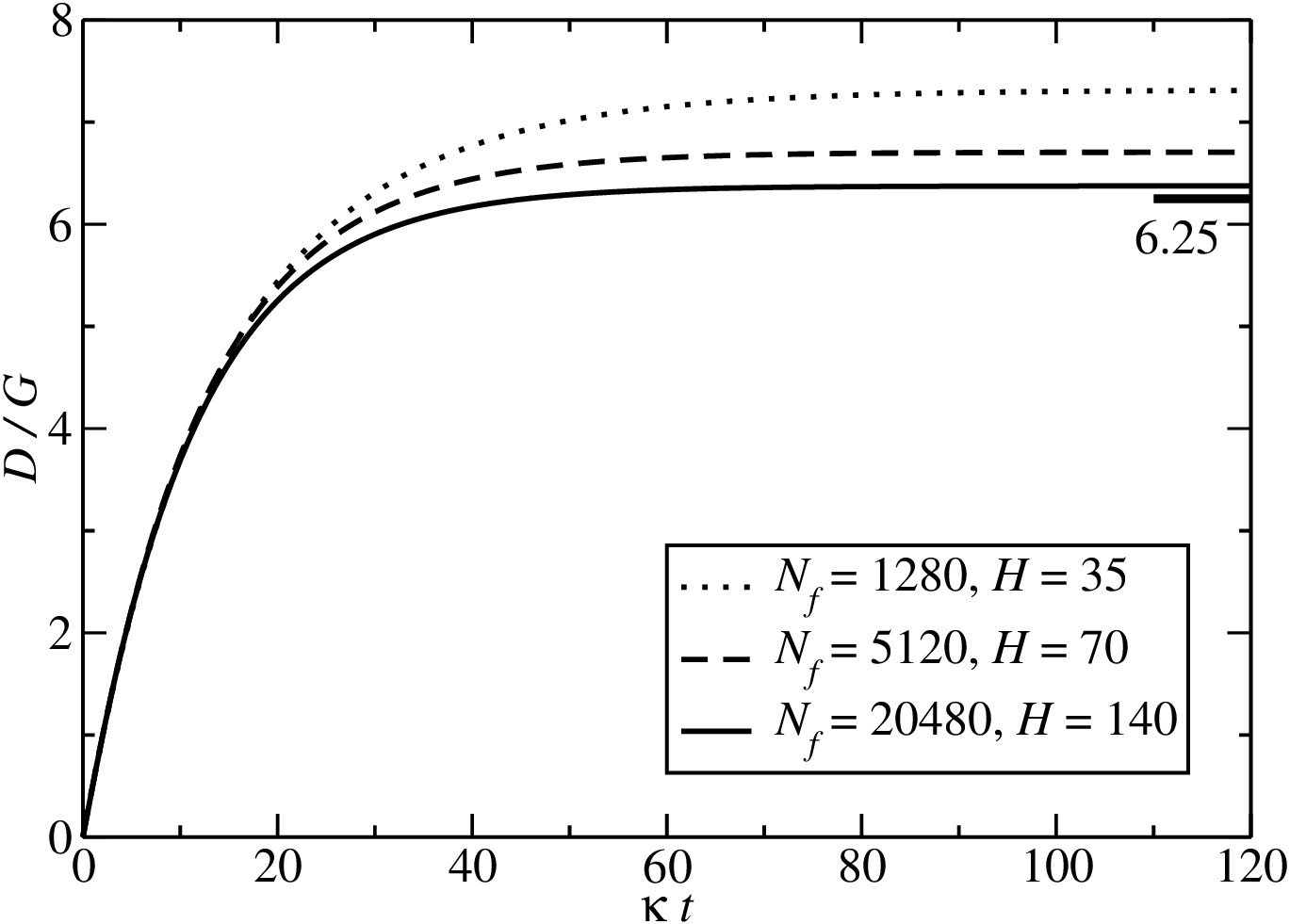}}
\subfigure[\label{fig:convergence_4_theta} inclination angle, $\phi_4$]{\includegraphics[width=0.45\linewidth,clip=true]{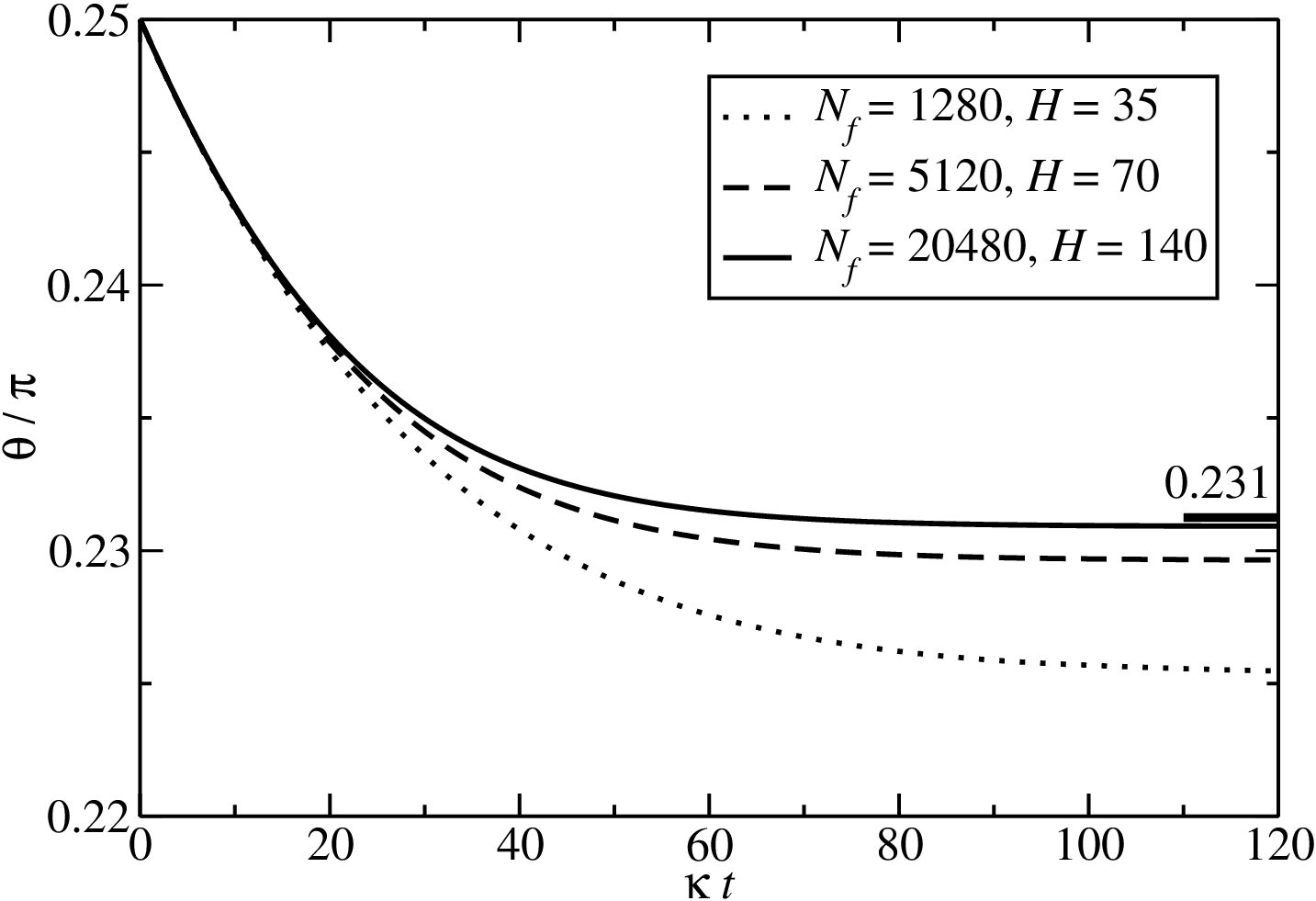}}
\caption{Convergence of the IBLBFEM for the interpolation stencils $\phi_2$ in subfigures \subref{fig:convergence_2_D} and \subref{fig:convergence_2_theta}, $\phi_3$ in subfigures \subref{fig:convergence_3_D} and \subref{fig:convergence_3_theta}, and $\phi_4$ in subfigures \subref{fig:convergence_4_D} and \subref{fig:convergence_4_theta}. The mesh ratio $\bar l / \Delta x$ is $0.53$ in all simulations. The mesh and fluid resolutions are $N_f = 1280$ and $H = 35$ (dotted lines), $N_f = 5120$ and $H = 70$ (dashed lines), and $N_f = 20480$ and $H = 140$ (solid lines), respectively. Convergence to the analytic predictions is observed in all cases. The numerical errors at $\kappa t = 120$ ($\kappa = \dot \gamma / G$) are also shown in Tab.\ \ref{tab:convergence} and Fig.\ \ref{fig:convergence_2}.}
\label{fig:convergence}
\end{figure*}

\begin{table*}
\begin{ruledtabular}
\centering
\caption{\label{tab:convergence} Convergence of the IBLBFEM for the interpolation stencils $\phi_2$, $\phi_3$, and $\phi_4$. The deviations of the deformation parameter $D$ and the inclination angle $\theta$ at $\kappa t = \dot \gamma t / G = 120$ are shown. The relative deviations are defined as $\delta D = (D_s - D_a) / D_a$ and $\delta \theta^o = (\theta^o_s - \theta^o_a) / \theta^o_a$. The subscripts $s$ and $a$ denote `simulation' and `analytic'. The angle $\theta^o \propto G$ is the opposite angle to $\theta$, defined in Eq.\ (\ref{eq:analyticangle}). The convergence order $\alpha$ is taken from a fit to the function $\delta D, \delta \theta^o \propto N_r^{-\alpha}$. For $\delta \theta^o$ and $\phi_2$, a meaningful convergence order could not be obtained due to early mesh degradation. A graphic representation of this table is shown in Fig.\ \ref{fig:convergence_2}.}
\begin{tabularx}{\linewidth}{rrrrrrrr}
& & \multicolumn{2}{c}{$\phi_2$} & \multicolumn{2}{c}{$\phi_3$} & \multicolumn{2}{c}{$\phi_4$} \\
$H$ & $N_f$ & $\delta D$ & $\delta \theta^o$ & $\delta D$ & $\delta \theta^o$ & $\delta D$ & $\delta \theta^o$ \\ \hline
 35 &  1280 & $13.2\%$ & $12.0\%$ & $13.5\%$ & $20.9\%$ & $17.0\%$ & $30.8\%$ \\
 70 &  5120 & $4.1\%$ & $4.5\%$ & $4.5\%$ & $4.9\%$ & $7.3\%$ & $8.5\%$ \\
140 & 20480 & $1.2\%$ & $3.0\%$ & $1.0\%$ & $0.9\%$ & $2.0\%$ & $1.7\%$ \\ \hline
\multicolumn{2}{c}{convergence order} & $1.7$ & N/A & $1.9$ & $2.2$ & $1.5$ & $2.1$
\end{tabularx}
\end{ruledtabular}
\end{table*}

\begin{figure*}
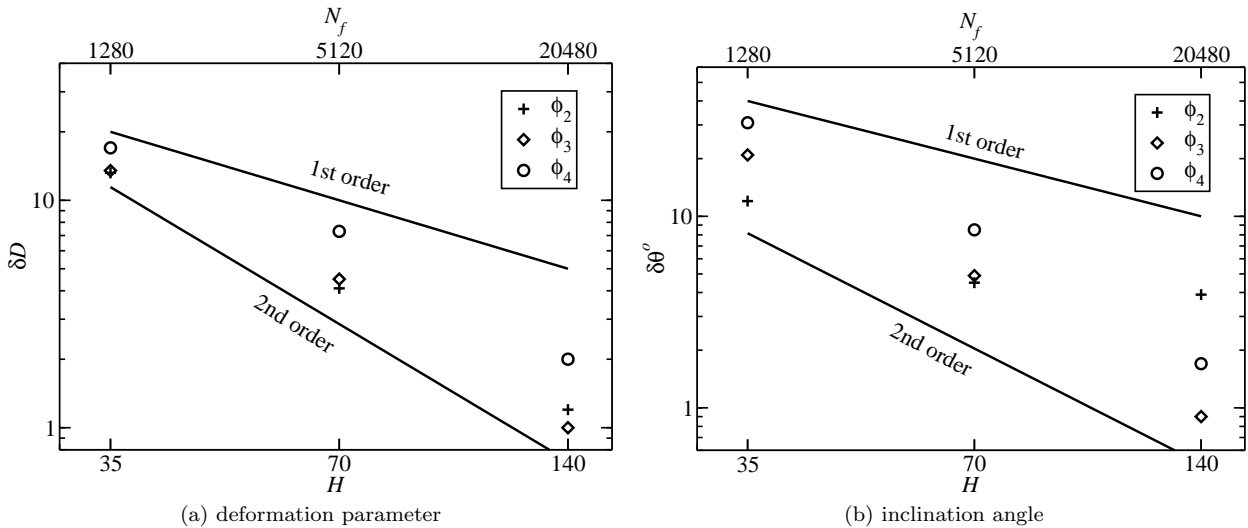

\centering
\subfigure[\label{fig:convergence_D} deformation parameter]{\includegraphics[width=0.45\linewidth,clip=true]{convergence_D}} \hspace{1ex}
\subfigure[\label{fig:convergence_theta} inclination angle]{\includegraphics[width=0.45\linewidth,clip=true]{convergence_theta}}
\caption{Convergence of the IBLBFEM for the interpolation stencils $\phi_2$, $\phi_3$, and $\phi_4$. In subfigure \subref{fig:convergence_D}, the error of the deformation parameter, $\delta D$, is shown for increasing mesh and fluid resolutions with fixed $\bar l / \Delta x = 0.53$. The analog results for the error of the inclination angle, $\delta \theta^o$, are shown in subfigure \subref{fig:convergence_theta}. The data is taken from Tab.\ \ref{tab:convergence}.}
\label{fig:convergence_2}
\end{figure*}

In this second series, we investigate the coupled convergence when both the mesh and the hydrodynamic resolutions are increased by the same rate, i.e., the mesh ratio $\bar l / \Delta x$ is fixed. The mesh and hydrodynamic resolutions are $N_f = 1280$ and $H = 35$, $N_f = 5120$ and $H = 70$, and $N_f = 20480$ and $H = 140$, respectively. Note that for $N_f = 320$, a mesh ratio of $\bar l / \Delta x = 0.53$ leads to quite inacceptable results. This is closely related to the fact that the hydrodynamic radius becomes too small compared to the numerical width of the membrane. Here, we see again that using the freedom in the choice of $\bar l / \Delta x$ allows us to use mesh resolutions which would not be available otherwise. The mesh ratio is $\bar l / \Delta x = 0.53$ in all cases. The results are shown in Fig.\ \ref{fig:convergence}. It is obvious that the steady-state magnitudes of $D$ and $\theta$ converge to their analytic values ($D_a/G = 6.25$ and $\theta_a/\pi \approx 0.231$). In order to quantify the results, the errors at $\kappa t = 120$ are listed in Tab.\ \ref{tab:convergence} and graphically shown in Fig.\ \ref{fig:convergence_2}. The convergence order is clearly better than $1$, cf.\ Tab.\ \ref{tab:convergence}. The only exception is the inclination angle with $\phi_2$. This is caused by mesh degradation, cf.\ Sec.\ \ref{subsec:degradation}. Since the LBM is second-order accurate and the IBM for sharp interfaces formally first order, the convergence order of the coupled system should be between $1$ and $2$ as observed here. Unfortunately, we are not aware of any theory which could predict the convergence behavior of the coupled system.

\paragraph{Effective deformability and rescaling}

From the results in this section, we draw two main observations.

First, the numerical magnitudes of the deformation parameter $D$ and the inclination angle $\theta$ approach the expected values when the hydrodynamic resolution $N_r$ is increased. Within the valid region of $\bar l / \Delta x$, this statement also holds if only the hydrodynamic resolution is increased and the mesh resolution is kept constant. However, the formally correct approach is to gradually refine both meshes simultaneously. The IBLBFEM accurately captures the physics of deformable capsules in an ambient fluid within the chosen parameter ranges and in the limit of infinite resolution. The convergence order is between $1.5$ and $2$, cf.\ Tab.\ \ref{tab:convergence} and Fig.\ \ref{fig:convergence_2}.

Second, we conclude that the deviations between expected and analytical values are an IBM artifact. The average deviations are usually smaller for a narrower interpolation stencil, cf.\ Tab.\ \ref{tab:convergence} (except $\phi_2$ at large resolution). This supports the idea that the numerical width of the interpolation stencil affects the deformation behavior of the capsule. The reason for the effect of $\phi_2$ at large resolutions is an accelerated mesh degradation. We will come back to this point in Sec.\ \ref{subsec:degradation}. On the first glance, a narrower interpolation stencil does a better job in capturing the physics of the problem. However, one finds that fluctuations are more pronounced when $\phi_3$ and especially $\phi_2$ are employed since those interpolations are not as smooth as $\phi_4$. This can be seen in the plots of $\theta$ in Figs.\ \ref{fig:ratio_320}, \ref{fig:ratio_1280}, and \ref{fig:convergence}. Making a good choice for the interpolation stencil means balancing the strengths and weaknesses of those stencils. We have also seen that, even for coarser resolutions, the numerical results are sound and reliable, as long as the presence of the finite membrane thickness is properly taken into account.

\begin{table*}
\begin{ruledtabular}
\centering
\caption{\label{tab:effshearrate} Effective reduced shear rates $\tilde G / G$ and effective errors $\delta \tilde \theta^o = (\theta^o_s - \theta^o_a(\tilde G)) / \theta^o_a(\tilde G)$ of the opposite angle. $\tilde G$ is defined in such a way that the deformation parameter has no error, i.e., $D_s = D_a(\tilde G)$. The effective deviations $\delta \tilde \theta^o$ are clearly reduced compared to the raw data in Tab.\ \ref{tab:convergence}.}
\begin{tabularx}{\linewidth}{rrrrrrrr}
& & \multicolumn{2}{c}{$\phi_2$} & \multicolumn{2}{c}{$\phi_3$} & \multicolumn{2}{c}{$\phi_4$} \\
$H$ & $N_f$ & $\tilde G / G$ & $\delta \tilde \theta^o$ & $\tilde G / G$ & $\delta \tilde \theta^o$ & $\tilde G / G$ & $\delta \tilde \theta^o$ \\ \hline
 35 &  1280 & $1.13$ & $-1.1\%$ & $1.14$ & $6.5\%$ & $1.17$ & $11.8\%$ \\
 70 &  5120 & $1.04$ & $0.4\%$ & $1.05$ & $0.4\%$ & $1.07$ & $1.1\%$ \\
140 & 20480 & $1.01$ & $1.8\%$ & $1.01$ & $-0.1\%$ & $1.02$ & $-0.3\%$
\end{tabularx}
\end{ruledtabular}
\end{table*}

It is worthwhile to discuss the deviations of the deformation parameter $D$ and the opposite inclination angle $\theta^o$, Eq.\ \ref{eq:analyticangle}, in more details. While the deviations $\delta D$ and $\delta \theta$ become smaller when the hydrodynamic resolution $N_r$ is increased, $\delta D$ and $\delta \theta^o$ are always of the same order. Since $D \propto G$ and $\theta^o \propto G$, this enables us to introduce an effective reduced shear rate $\tilde G$ to partially counteract the effect of the finite membrane thickness due to the presence of the interpolation stencil. Starting from Eq.\ (\ref{eq:shearrate}), the effective reduced shear rate can be written as
\be
\label{eq:effectiveshearrate}
\tilde G = \f{\dot \gamma \eta \tilde r}{\tilde k_s},
\ee
i.e., the transition $G \to \tilde G$ can be due to a transition $r \to \tilde r$ or $k_s \to \tilde k_s$. We claim that the fluid properties $\dot \gamma$ and $\eta$ are well defined, thus, we do not allow their effective modification. In the present simulations, we have the choice to define an effective radius $\tilde r$ or an effective stiffness $\tilde k_s$ in such a way that the simulation data are more accurate. It is left for future research whether a redefinition of $r$ or $k_s$ is more useful. In the present simulations, both approaches are equivalent, but due to particle-particle interactions in simulations with multiple capsules, the effect of the finite membrane width is more complex and a simple redefinition of $r$ or $k_s$ may be not straightforward. In general, the rescaling factor $\tilde G / G$ is a function of the interpolation stencil $\phi$ and the hydrodynamic and the mesh resolutions, $N_r$ and $N_f$. The factor $\tilde G / G$ can be obtained from Tab.\ \ref{tab:convergence}. In order to investigate this idea further, we have defined $\tilde G$ for each simulation in Tab.\ \ref{tab:convergence} in such a way that the deviation $\delta \tilde D := (D_s - D_a(\tilde G)) / D_a(\tilde G)$ is identically zero. The corresponding error $\delta \tilde \theta^o := (\theta^o_s - \theta^o_a (\tilde G)) / \theta^o_a (\tilde G)$ is shown in Tab.\ \ref{tab:effshearrate}. All effective errors $\delta \tilde \theta^o$ are considerably smaller than the original errors given in Tab.\ \ref{tab:convergence}, stating that the redefinition of $G$ can be used to compensate---at least partially---the numerical width effect due to the interpolation stencil. It is expected that for interacting capsules with more complex shape, e.g., red blood cells, a simple rescaling is not as straightforward as for an isolated spherical capsule.

To sum up the above findings, we note that the dominant source of numerical deviation from the analytic solution in the combined IBLBFEM is the presence of the interpolation stencils. They introduce a numerical width of the membrane, leading to slightly modified physical behavior. Those deviations could in principle be counteracted by introducing an effective radius or effective stiffness of the capsule. In the limit of infinite resolution, the IBLFFEM accurately captures the physics of the coupled system of capsule and fluid. The influence of the choice of $\bar l / \Delta x$ is small over a wide range, $0.5 < \bar l / \Delta x < 1.5$. This introduces a freedom to the numerical implementation which can be used to decrease the computational cost of the simulations.

\subsection{Mesh degradation and volume drift}
\label{subsec:degradation}

\begin{figure*}
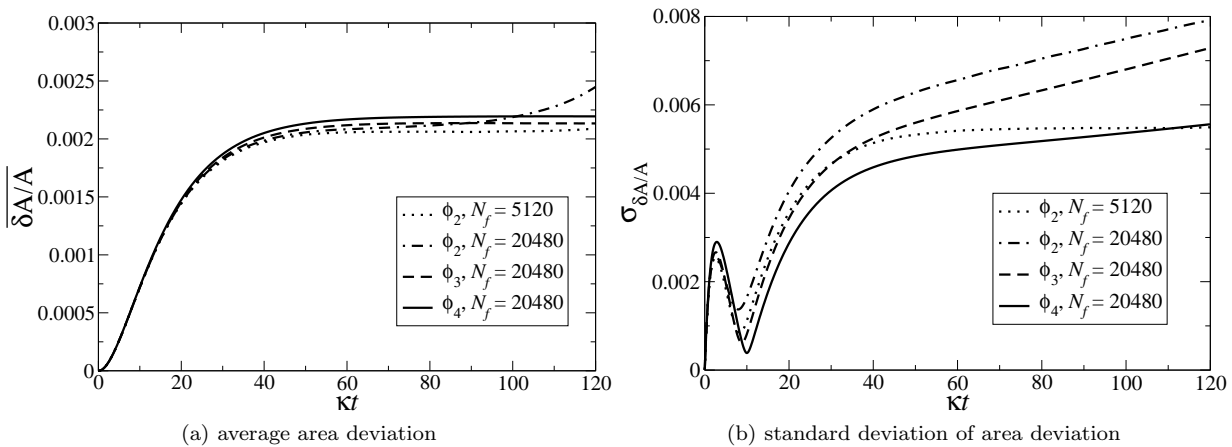

\centering
\subfigure[\label{fig:degrademean} average area deviation]{\includegraphics[width=0.45\linewidth,clip=true]{degrade_mean}}
\subfigure[\label{fig:degradesigma} standard deviation of area deviation]{\includegraphics[width=0.45\linewidth,clip=true]{degrade_sigma}}
\caption{Time evolution of average and standard deviation of the relative face area deviation $\delta A / A$. At $t = 0$, all faces are undeformed and hence $\overline{\delta A / A} = 0$ and $\sigma_{\delta A / A} = 0$. Due to the deformation of the capsule, \subref{fig:degrademean} the average area deviation $\overline{\delta A / A}$ increases until it reaches a steady state. The steady state eventually collapses when the mesh degrades. This is most significant for $\phi_2$ at the highest resolution of the membrane mesh ($N_f = 20480$). The degradation is much less significant for $\phi_2$ and $N_f = 5120$, indicating that too fine a mesh (i.e., too coarse a hydrodynamic resolution) is detrimental. \subref{fig:degradesigma} The standard deviation $\sigma_{\delta A / A}$ of the relative area deviation increases with time, especially for a mesh with high resolution but poor hydrodynamic resolution.}
\label{fig:degrade}
\end{figure*}

In this section, we will address some issues which have not been covered in the discussion about the mesh influence and the convergence studies in Secs.\ \ref{subsec:finite_size}--\ref{subsec:interpolation}.

Although we have claimed that the mesh resolution does not have a significant influence on the results (cf.\ Sec.\ \ref{subsec:meshdiscretization}), we have reported different deviations of the deformation parameter $D$ and the inclination angle $\theta$ when $\phi_2$ and varying mesh resolutions are used (cf.\ Tab.\ \ref{tab:convergence}). The reason is that the values in Tab.\ \ref{tab:convergence} are taken at time $\kappa t = 120$ where the mesh has already started to degrade when $\phi_2$ is employed. This effect seems to be most severe when the hydrodynamic resolution is smallest (i.e., small radius $N_r$ and large number of faces $N_f$). A similar behavior is only weakly noticeable for $\phi_3$ and $\phi_4$, indicating that those interpolation stencils do a better job in preserving the mesh. This observation is an indication that the average node distance $\bar l / \Delta x$ should not be too small. The mesh degradation can be captured by computing the average face element area and its standard deviation. For a degrading mesh, one expects those values to leave the steady state gradually. Our findings are illustrated in Fig.\ \ref{fig:degrade}. Only a few curves are shown. The major observation is that increasing the mesh resolution but keeping the hydrodynamic resolution unchanged leads to an accelerated and undesired mesh degradation. Although the physical behavior of the capsule is independent of the mesh resolution at small times (cf.\ Sec.\ \ref{subsec:meshdiscretization}), numerical artifacts become progressively more important at later times. For long-time simulations, the shear energy $W^S$ may be not sufficient to control the mesh.

Taking the results of the simulations presented in Sec.\ \ref{subsec:interpolation}, we have observed typical volume deviations $\delta V / V_0$ between $2 \cdot 10^{-4}$ and $8 \cdot 10^{-4}$ for $\phi_2$, between $5 \cdot 10^{-5}$ and $2 \cdot 10^{-4}$ for $\phi_3$, and between $9 \cdot 10^{-6}$ and $3 \cdot 10^{-5}$ for $\phi_4$ at $\kappa t = 120$. As expected, the largest deviations correspond to the smallest radius $N_r$ and vice versa. Clearly, the 4-point interpolation function is much more appropriate to control the capsule volume without taking additional measures. A higher hydrodynamic resolution also reduces the volume drift.

In the present simulations, the volume drift is negligible. However, in long-time simulations with smaller resolution, the volume drift could become a significant problem. Taking into account restoring forces originating from volume, surface, and bending energies may help to avoid this deficiency.


\section{Conclusions}
\label{sec:conclusion}

We found that the choice of the LBM relaxation parameter $\tau$ can have a detrimental effect on the combined IBM-LBM simulations. As long as $\tau \leq 1$, the simulation results barely depend on the actual magnitude of $\tau$. For larger relaxation parameters, strong deviations can be observed, and the simulation results obtained by different $\tau$ are no longer comparable. On the other hand, we have seen that the choice of $\tau$ strongly affects the simulation time since $\tau$ can be used to change the LBM time step. We have used $\tau = 1$ in the simulations as a compromise between accuracy and efficiency.

The common assumption that the average node distance $\bar l$ should be smaller than half a fluid lattice constant $\Delta x$ could not be supported in our investigations. This leads to the important consequence that, at least within a certain range, the hydrodynamic resolution and the mesh resolution can be changed independently without significantly changing the physics in the simulations.

Due to its locality and purely algebraic structure, the LBM algorithm is fast and efficient. It is well known that the computing speed of pure LBM simulations is usually restricted by memory access, but not by CPU power. Considering the capsule immersed in the fluid, the computation of the membrane forces, the velocity interpolation, and the force spreading is computationally more demanding and basically limited by the CPU power. In the present simulations, the fluid volume is large compared to the membrane area, and nearly all the computing time is consumed by the LBM component. For simulations with moderate or high particle volume fractions, however, the IBM consumes most of the computer resources. We have observed that the computing time for the velocity interpolation and spreading of the forces is much larger than that for the computation of the membrane forces itself. The number of IBM interpolations at each time step is $2 m^d N_n$ in $d$ dimensions, using a stencil $\phi_m$ with a support of $m$ lattice nodes along each direction and a total of $N_n$ membrane nodes. One way to save computing time is to choose an interpolation stencil with smaller support. Although the relevant physics seems to be captured with the 2-point stencil as well, stronger fluctuations are introduced, and the mesh undergoes an accelerated degradation (cf.\ Secs.\ \ref{subsec:interpolation} and \ref{subsec:degradation}). However, a smaller support is also equivalent to a smaller numerical thickness of the membranes, which could be of advantage in a dense suspension of deformable particles where the distances between the membranes are small.

The major discretization error is introduced by the IBM interpolation stencil $\phi$, but not by the discretization of the membrane itself. The hydrodynamic resolution is more important than the mesh resolution. This observation leads to the following conclusion: if the accuracy of the simulation shall be increased, it is the optimum approach to first increase the number of fluid lattice nodes for fixed mesh resolution. If on the other hand the computing time of the simulation should be decreased, the hydrodynamic resolution should be kept and the mesh resolution be reduced.

We observed that the presence of the interpolation stencils $\phi$ effectively changes the membrane properties of the capsules due to the numerical thickness of the membrane. This can be captured by defining an effective reduced shear rate $\tilde G$ by considering an effective capsule radius $\tilde r$ or an effective stiffness $\tilde k_s$. The numerical membrane width decreases when the hydrodynamic resolution is refined, i.e., when the number $N_r$ of fluid lattice nodes covered by the capsule radius is increased. The numerical results converge to the analytically expected values for large $N_r$. Consequently, making use of the knowledge of an effective radius $\tilde r$ or stiffness $\tilde k_s$ in principle allows for the reduction of the hydrodynamic resolution without a significant loss of accuracy. This approach may be used to massively save computing time in simulations of multiple deformable particles immersed in a fluid.

We stress that, in this paper, only small deformations of a single capsule have been considered. The mesh resolution plays only a minor role in this case. However, the number of mesh points may be of higher importance in simulations with large membrane deformations since the faces remain always flat. For that reason, the analysis of the influence of the mesh resolution should also be performed for large deformations. The effect of particle-particle interactions in dense suspensions on an effective radius or stiffness caused by the interpolation stencils is also not obvious at this point. Those investigations are left for future research.

The present paper contains new contributions: regarding the effect of the interpolation stencils on the capsules' deformation behavior, the significance of the hydrodynamic resolution compared to the mesh resolution, and how those insights may be used to boost the efficiency of the related simulations. These results are hoped to be useful for the simulation of multiple deformable objects immersed in a fluid.


\begin{acknowledgements}

This project has been supported by the DFG grant VA205/5-1. We acknowledge fruitful discussions with Markus Gross, Denny Tjahjanto, Hong Tong Low, Georgios Zikos, Dennis Hessling, and Segun Ayodele.

\end{acknowledgements}


\small


\begin{thebibliography}{64}
\providecommand{\natexlab}[1]{#1}
\providecommand{\url}[1]{\texttt{#1}}
\providecommand{\urlprefix}{URL }
\expandafter\ifx\csname urlstyle\endcsname\relax
  \providecommand{\doi}[1]{doi:\discretionary{}{}{}#1}\else
  \providecommand{\doi}[1]{doi:\discretionary{}{}{}\begingroup
  \urlstyle{rm}\url{#1}\endgroup}\fi
\providecommand{\bibinfo}[2]{#2}

\bibitem[{Barthès-Biesel(1980)}]{barths-biesel_motion_1980}
\bibinfo{author}{D.~Barthès-Biesel}, \bibinfo{title}{Motion of a Spherical
  Microcapsule Freely Suspended in a Linear Shear Flow}, \bibinfo{journal}{J.
  Fluid Mech.} \bibinfo{volume}{100}~(\bibinfo{number}{04})
  (\bibinfo{year}{1980}) \bibinfo{pages}{831--853}.

\bibitem[{Barthès-Biesel and
  Rallison(1981)}]{barths-biesel_time-dependent_1981}
\bibinfo{author}{D.~Barthès-Biesel}, \bibinfo{author}{J.~M. Rallison},
  \bibinfo{title}{The Time-Dependent Deformation of a Capsule Freely Suspended
  in a Linear Shear Flow}, \bibinfo{journal}{J. Fluid Mech.}
  \bibinfo{volume}{113} (\bibinfo{year}{1981}) \bibinfo{pages}{251--267}.

\bibitem[{Chang and Olbricht(1993)}]{chang_experimental_1993}
\bibinfo{author}{K.~S. Chang}, \bibinfo{author}{W.~L. Olbricht},
  \bibinfo{title}{Experimental Studies of the Deformation and Breakup of a
  Synthetic Capsule in Steady and Unsteady Simple Shear Flow},
  \bibinfo{journal}{J. Fluid Mech.} \bibinfo{volume}{250}
  (\bibinfo{year}{1993}) \bibinfo{pages}{609--633}.

\bibitem[{Walter et~al.(2000)Walter, Rehage, and
  Leonhard}]{walter_shear-induced_2000}
\bibinfo{author}{A.~Walter}, \bibinfo{author}{H.~Rehage},
  \bibinfo{author}{H.~Leonhard}, \bibinfo{title}{Shear-induced deformations of
  polyamide microcapsules}, \bibinfo{journal}{Colloid Polym Sci}
  \bibinfo{volume}{278}~(\bibinfo{number}{2}) (\bibinfo{year}{2000})
  \bibinfo{pages}{169--175}.

\bibitem[{Pozrikidis(1995)}]{pozrikidis_finite_1995}
\bibinfo{author}{C.~Pozrikidis}, \bibinfo{title}{Finite Deformation of Liquid
  Capsules Enclosed by Elastic Membranes in Simple Shear Flow},
  \bibinfo{journal}{J. Fluid Mech.} \bibinfo{volume}{297}
  (\bibinfo{year}{1995}) \bibinfo{pages}{123--152}.

\bibitem[{Kraus et~al.(1996)Kraus, Wintz, Seifert, and
  Lipowsky}]{kraus_fluid_1996}
\bibinfo{author}{M.~Kraus}, \bibinfo{author}{W.~Wintz},
  \bibinfo{author}{U.~Seifert}, \bibinfo{author}{R.~Lipowsky},
  \bibinfo{title}{Fluid Vesicles in Shear Flow}, \bibinfo{journal}{Phys. Rev.
  Lett.} \bibinfo{volume}{77}~(\bibinfo{number}{17}) (\bibinfo{year}{1996})
  \bibinfo{pages}{3685}.

\bibitem[{Ramanujan and Pozrikidis(1998)}]{ramanujan_deformation_1998}
\bibinfo{author}{S.~Ramanujan}, \bibinfo{author}{C.~Pozrikidis},
  \bibinfo{title}{Deformation of Liquid Capsules Enclosed by Elastic Membranes
  in Simple Shear Flow: Large Deformations and the Effect of Fluid
  Viscosities}, \bibinfo{journal}{J. Fluid Mech.} \bibinfo{volume}{361}
  (\bibinfo{year}{1998}) \bibinfo{pages}{117--143}.

\bibitem[{Diaz et~al.(2000)Diaz, Pelekasis, and
  {Barthes-Biesel}}]{diaz_transient_2000}
\bibinfo{author}{A.~Diaz}, \bibinfo{author}{N.~Pelekasis},
  \bibinfo{author}{D.~{Barthes-Biesel}}, \bibinfo{title}{Transient response of
  a capsule subjected to varying flow conditions: Effect of internal fluid
  viscosity and membrane elasticity}, \bibinfo{journal}{Phys. Fluids}
  \bibinfo{volume}{12}~(\bibinfo{number}{5}) (\bibinfo{year}{2000})
  \bibinfo{pages}{948--957}.

\bibitem[{Barthès-Biesel et~al.(2002)Barthès-Biesel, Diaz, and
  Dhenin}]{barths-biesel_effect_2002}
\bibinfo{author}{D.~Barthès-Biesel}, \bibinfo{author}{A.~Diaz},
  \bibinfo{author}{E.~Dhenin}, \bibinfo{title}{Effect of constitutive laws for
  two-dimensional membranes on flow-induced capsule deformation},
  \bibinfo{journal}{J. Fluid Mech.} \bibinfo{volume}{460}
  (\bibinfo{year}{2002}) \bibinfo{pages}{211--222}.

\bibitem[{Lac et~al.(2004)Lac, Barthès-Biesel, Pelekasis, and
  Tsamopoulos}]{lac_spherical_2004}
\bibinfo{author}{E.~Lac}, \bibinfo{author}{D.~Barthès-Biesel},
  \bibinfo{author}{N.~A. Pelekasis}, \bibinfo{author}{J.~Tsamopoulos},
  \bibinfo{title}{Spherical capsules in three-dimensional unbounded Stokes
  flows: effect of the membrane constitutive law and onset of buckling},
  \bibinfo{journal}{J. Fluid Mech.} \bibinfo{volume}{516}
  (\bibinfo{year}{2004}) \bibinfo{pages}{303--334}.

\bibitem[{Eggleton and Popel(1998)}]{eggleton_large_1998}
\bibinfo{author}{C.~D. Eggleton}, \bibinfo{author}{A.~S. Popel},
  \bibinfo{title}{Large Deformation of Red Blood Cell Ghosts in a Simple Shear
  Flow}, \bibinfo{journal}{Phys. Fluids}
  \bibinfo{volume}{10}~(\bibinfo{number}{8}) (\bibinfo{year}{1998})
  \bibinfo{pages}{1834--1845}.

\bibitem[{Sui et~al.(2008)Sui, Chew, Roy, and Low}]{sui_hybrid_2008}
\bibinfo{author}{Y.~Sui}, \bibinfo{author}{Y.~T. Chew},
  \bibinfo{author}{P.~Roy}, \bibinfo{author}{H.~T. Low}, \bibinfo{title}{A
  hybrid method to study flow-induced deformation of three-dimensional
  capsules}, \bibinfo{journal}{J. Comput. Phys.}
  \bibinfo{volume}{227}~(\bibinfo{number}{12}) (\bibinfo{year}{2008})
  \bibinfo{pages}{6351--6371}.

\bibitem[{Pozrikidis(2001)}]{pozrikidis_effect_2001}
\bibinfo{author}{C.~Pozrikidis}, \bibinfo{title}{Effect of Membrane Bending
  Stiffness on the Deformation of Capsules in Simple Shear Flow},
  \bibinfo{journal}{J. Fluid Mech.} \bibinfo{volume}{440}
  (\bibinfo{year}{2001}) \bibinfo{pages}{269--291}.

\bibitem[{Dupin et~al.(2007)Dupin, Halliday, Care, Alboul, and
  Munn}]{dupin_modelingflow_2007}
\bibinfo{author}{M.~M. Dupin}, \bibinfo{author}{I.~Halliday},
  \bibinfo{author}{C.~M. Care}, \bibinfo{author}{L.~Alboul},
  \bibinfo{author}{L.~L. Munn}, \bibinfo{title}{Modeling the Flow of Dense
  Suspensions of Deformable Particles in Three Dimensions},
  \bibinfo{journal}{Phys. Rev. E} \bibinfo{volume}{75}~(\bibinfo{number}{6})
  (\bibinfo{year}{2007}) \bibinfo{pages}{066707--17}.

\bibitem[{Doddi and Bagchi(2009)}]{doddi_three-dimensional_2009}
\bibinfo{author}{S.~K. Doddi}, \bibinfo{author}{P.~Bagchi},
  \bibinfo{title}{Three-dimensional computational modeling of multiple
  deformable cells flowing in microvessels}, \bibinfo{journal}{Phys. Rev. E}
  \bibinfo{volume}{79}~(\bibinfo{number}{4}) (\bibinfo{year}{2009})
  \bibinfo{pages}{046318}.

\bibitem[{Qian et~al.(1992)Qian, D'Humières, and
  Lallemand}]{qian_lattice_1992}
\bibinfo{author}{Y.~H. Qian}, \bibinfo{author}{D.~D'Humières},
  \bibinfo{author}{P.~Lallemand}, \bibinfo{title}{Lattice BGK Models for
  Navier-Stokes Equation}, \bibinfo{journal}{Europhys. Lett.}
  \bibinfo{volume}{17} (\bibinfo{year}{1992}) \bibinfo{pages}{479}.

\bibitem[{Chen and Doolen(1998)}]{chen_lattice_1998}
\bibinfo{author}{S.~Chen}, \bibinfo{author}{G.~D. Doolen},
  \bibinfo{title}{Lattice Boltzmann Method for Fluid Flows},
  \bibinfo{journal}{Annu. Rev. Fluid Mech.} \bibinfo{volume}{30}
  (\bibinfo{year}{1998}) \bibinfo{pages}{329--364}.

\bibitem[{Ladd and Verberg(2001)}]{ladd2001lbs}
\bibinfo{author}{A.~J.~C. Ladd}, \bibinfo{author}{R.~Verberg},
  \bibinfo{title}{Lattice-Boltzmann simulations of particle-fluid suspensions},
  \bibinfo{journal}{J. Stat. Phys.} \bibinfo{volume}{104}~(\bibinfo{number}{5})
  (\bibinfo{year}{2001}) \bibinfo{pages}{1191--1251}.

\bibitem[{Succi(2001)}]{succi_lattice_2001}
\bibinfo{author}{S.~Succi}, \bibinfo{title}{The Lattice Boltzmann Equation for
  Fluid Dynamics and Beyond}, \bibinfo{publisher}{Oxford University Press,
  USA}, ISBN \bibinfo{isbn}{0198503989}, \bibinfo{year}{2001}.

\bibitem[{Sukop and Thorne(2005)}]{sukop_lattice_2005}
\bibinfo{author}{M.~Sukop}, \bibinfo{author}{D.~Thorne},
  \bibinfo{title}{Lattice Boltzmann Modeling, an Introduction for Geoscientists
  and Engineers}, \bibinfo{publisher}{Springer}, \bibinfo{address}{Berlin
  Heidelberg}, ISBN \bibinfo{isbn}{978-3-540-27981-5}, \bibinfo{year}{2005}.

\bibitem[{Ayodele et~al.(2009)Ayodele, Varnik, and Raabe}]{ayodele_effect_2009}
\bibinfo{author}{S.~G. Ayodele}, \bibinfo{author}{F.~Varnik},
  \bibinfo{author}{D.~Raabe}, \bibinfo{title}{Effect of aspect ratio on
  transverse diffusive broadening: A lattice Boltzmann study},
  \bibinfo{journal}{Phys. Rev. E} \bibinfo{volume}{80}~(\bibinfo{number}{1})
  (\bibinfo{year}{2009}) \bibinfo{pages}{016304--9}.

\bibitem[{Varnik et~al.(2007)Varnik, Dorner, and
  Raabe}]{varnik_roughness-induced_2007}
\bibinfo{author}{F.~Varnik}, \bibinfo{author}{D.~Dorner},
  \bibinfo{author}{D.~Raabe}, \bibinfo{title}{{Roughness-Induced} Flow
  Instability: A Lattice Boltzmann Study}, \bibinfo{journal}{J. Fluid Mech.}
  \bibinfo{volume}{573} (\bibinfo{year}{2007}) \bibinfo{pages}{191--209}.

\bibitem[{Varnik et~al.(2008)Varnik, Truman, Wu, Uhlmann, Raabe, and
  Stamm}]{varnik_wetting_2008}
\bibinfo{author}{F.~Varnik}, \bibinfo{author}{P.~Truman},
  \bibinfo{author}{B.~Wu}, \bibinfo{author}{P.~Uhlmann},
  \bibinfo{author}{D.~Raabe}, \bibinfo{author}{M.~Stamm},
  \bibinfo{title}{Wetting gradient induced separation of emulsions: A combined
  experimental and lattice Boltzmann computer simulation study},
  \bibinfo{journal}{Phys. Fluid} \bibinfo{volume}{20}~(\bibinfo{number}{7})
  (\bibinfo{year}{2008}) \bibinfo{pages}{072104--14}.

\bibitem[{Gross et~al.(2009)Gross, Varnik, and Raabe}]{markus_acc}
\bibinfo{author}{M.~Gross}, \bibinfo{author}{F.~Varnik},
  \bibinfo{author}{D.~Raabe}, \bibinfo{title}{Fall and rise of small droplets
  on rough hydrophobic substrates},
  \bibinfo{journal}{Europhys. Lett.} \bibinfo{volume}{88}
  (\bibinfo{year}{2009}) \bibinfo{pages}{26002}.

\bibitem[{Peskin(1972)}]{peskin1972fpa}
\bibinfo{author}{C.~S. Peskin}, \bibinfo{title}{{Flow patterns around heart
  valves: A digital computer method for solving the equations of motion}},
  \bibinfo{publisher}{Sue Golding Graduate Division of Medical Sciences, Albert
  Einstein College of Medicine, Yeshiva University}, \bibinfo{year}{1972}.

\bibitem[{Peskin(2002)}]{peskin_ibm_2002}
\bibinfo{author}{C.~S. Peskin}, \bibinfo{title}{The Immersed Boundary Method},
  \bibinfo{journal}{Acta Numerica}  (\bibinfo{year}{2002})
  \bibinfo{pages}{479--517}.

\bibitem[{Feng and Michaelides(2004)}]{feng_immersed_2004}
\bibinfo{author}{Z.-G. Feng}, \bibinfo{author}{E.~E. Michaelides},
  \bibinfo{title}{The Immersed Boundary-Lattice Boltzmann Method for Solving
  Fluid-Particles Interaction Problems}, \bibinfo{journal}{J. Comput. Phys.}
  \bibinfo{volume}{195}~(\bibinfo{number}{2}) (\bibinfo{year}{2004})
  \bibinfo{pages}{602--628}.

\bibitem[{Niu et~al.(2006)Niu, Shu, Chew, and Peng}]{niu_momentum_2006}
\bibinfo{author}{X.~D. Niu}, \bibinfo{author}{C.~Shu}, \bibinfo{author}{Y.~T.
  Chew}, \bibinfo{author}{Y.~Peng}, \bibinfo{title}{A momentum exchange-based
  immersed boundary-lattice Boltzmann method for simulating incompressible
  viscous flows}, \bibinfo{journal}{Phys. Lett. A}
  \bibinfo{volume}{354}~(\bibinfo{number}{3}) (\bibinfo{year}{2006})
  \bibinfo{pages}{173--182}.

\bibitem[{Strack and Cook(2007)}]{strack_three-dimensional_2007}
\bibinfo{author}{O.~E. Strack}, \bibinfo{author}{B.~K. Cook},
  \bibinfo{title}{Three-dimensional immersed boundary conditions for moving
  solids in the {lattice-Boltzmann} method}, \bibinfo{journal}{Int. J. Numer.
  Meth. Fluids} \bibinfo{volume}{55} (\bibinfo{year}{2007})
  \bibinfo{pages}{103--125}.

\bibitem[{Peng and Luo(2008)}]{peng_comparative_2008}
\bibinfo{author}{Y.~Peng}, \bibinfo{author}{L.-S. Luo}, \bibinfo{title}{A
  comparative study of immersed-boundary and interpolated bounce-back methods
  in {LBE}}, \bibinfo{journal}{Progress in Computational Fluid Dynamics}
  \bibinfo{volume}{8}~(\bibinfo{number}{1/2/3/4}) (\bibinfo{year}{2008})
  \bibinfo{pages}{156 -- 167}.

\bibitem[{Ginzbourg and d'Humières(1996)}]{ginzbourg_local_1996}
\bibinfo{author}{I.~Ginzbourg}, \bibinfo{author}{D.~d'Humières},
  \bibinfo{title}{Local second-order boundary methods for lattice Boltzmann
  models}, \bibinfo{journal}{J. Stat. Phys.}
  \bibinfo{volume}{84}~(\bibinfo{number}{5}) (\bibinfo{year}{1996})
  \bibinfo{pages}{927--971}.

\bibitem[{Lallemand and Luo(2003)}]{lallemand_lattice_2003}
\bibinfo{author}{P.~Lallemand}, \bibinfo{author}{L.-S. Luo},
  \bibinfo{title}{Lattice Boltzmann method for moving boundaries},
  \bibinfo{journal}{J. Comput. Phys.}
  \bibinfo{volume}{184}~(\bibinfo{number}{2}) (\bibinfo{year}{2003})
  \bibinfo{pages}{406--421}.

\bibitem[{Ding and Aidun(2003)}]{ding_extension_2003}
\bibinfo{author}{E.-J. Ding}, \bibinfo{author}{C.~K. Aidun},
  \bibinfo{title}{Extension of the {Lattice-Boltzmann} Method for Direct
  Simulation of Suspended Particles Near Contact}, \bibinfo{journal}{J. Stat.
  Phys.} \bibinfo{volume}{112}~(\bibinfo{number}{3}) (\bibinfo{year}{2003})
  \bibinfo{pages}{685--708}.

\bibitem[{Chun and Ladd(2007)}]{chun_interpolated_2007}
\bibinfo{author}{B.~Chun}, \bibinfo{author}{A.~J.~C. Ladd},
  \bibinfo{title}{Interpolated boundary condition for lattice Boltzmann
  simulations of flows in narrow gaps}, \bibinfo{journal}{Phys. Rev. E}
  \bibinfo{volume}{75}~(\bibinfo{number}{6}) (\bibinfo{year}{2007})
  \bibinfo{pages}{066705--12}.

\bibitem[{Zhang et~al.(2007)Zhang, Johnson, and Popel}]{zhang_immersed_2007}
\bibinfo{author}{J.~Zhang}, \bibinfo{author}{P.~C. Johnson},
  \bibinfo{author}{A.~S. Popel}, \bibinfo{title}{An Immersed Boundary Lattice
  Boltzmann Approach to Simulate Deformable Liquid Capsules and its Application
  to Microscopic Blood Flows}, \bibinfo{journal}{Phys. Biol.}
  \bibinfo{volume}{4}~(\bibinfo{number}{4}) (\bibinfo{year}{2007})
  \bibinfo{pages}{285--295}.

\bibitem[{Peskin and Printz(1993)}]{peskin_improved_1993}
\bibinfo{author}{C.~S. Peskin}, \bibinfo{author}{B.~F. Printz},
  \bibinfo{title}{Improved Volume Conservation in the Computation of Flows with
  Immersed Elastic Boundaries}, \bibinfo{journal}{J. Comput. Phys.}
  \bibinfo{volume}{105} (\bibinfo{year}{1993}) \bibinfo{pages}{33--46}.

\bibitem[{Newren et~al.(2007)Newren, Fogelson, Guy, and
  Kirby}]{newren_unconditionally_2007}
\bibinfo{author}{E.~P. Newren}, \bibinfo{author}{A.~L. Fogelson},
  \bibinfo{author}{R.~D. Guy}, \bibinfo{author}{R.~M. Kirby},
  \bibinfo{title}{Unconditionally stable discretizations of the immersed
  boundary equations}, \bibinfo{journal}{J. Comput. Phys.}
  \bibinfo{volume}{222}~(\bibinfo{number}{2}) (\bibinfo{year}{2007})
  \bibinfo{pages}{702--719}.

\bibitem[{Shu et~al.(2007)Shu, Liu, and Chew}]{shu_novel_2007}
\bibinfo{author}{C.~Shu}, \bibinfo{author}{N.~Liu}, \bibinfo{author}{Y.~T.
  Chew}, \bibinfo{title}{A novel immersed boundary velocity correction-lattice
  Boltzmann method and its application to simulate flow past a circular
  cylinder}, \bibinfo{journal}{J. Comput. Phys.}
  \bibinfo{volume}{226}~(\bibinfo{number}{2}) (\bibinfo{year}{2007})
  \bibinfo{pages}{1607--1622}.

\bibitem[{Wu and Shu(2009)}]{wu_implicit_2009}
\bibinfo{author}{J.~Wu}, \bibinfo{author}{C.~Shu}, \bibinfo{title}{Implicit
  velocity correction-based immersed boundary-lattice Boltzmann method and its
  applications}, \bibinfo{journal}{Journal of Computational Physics}
  \bibinfo{volume}{228}~(\bibinfo{number}{6}) (\bibinfo{year}{2009})
  \bibinfo{pages}{1963--1979}.

\bibitem[{Tu and Peskin(1992)}]{tu_stability_1992}
\bibinfo{author}{C.~Tu}, \bibinfo{author}{C.~S. Peskin},
  \bibinfo{title}{Stability and Instability in the Computation of Flows with
  Moving Immersed Boundaries: A Comparison of Three Methods},
  \bibinfo{journal}{SIAM J. Sci. Stat. Comp.}
  \bibinfo{volume}{13}~(\bibinfo{number}{6}) (\bibinfo{year}{1992})
  \bibinfo{pages}{1361--1376}.

\bibitem[{Mori and Peskin(2008)}]{mori_implicit_2008}
\bibinfo{author}{Y.~Mori}, \bibinfo{author}{C.~S. Peskin},
  \bibinfo{title}{Implicit second-order immersed boundary methods with boundary
  mass}, \bibinfo{journal}{Comput. Method Appl. M.}
  \bibinfo{volume}{197}~(\bibinfo{number}{25-28}) (\bibinfo{year}{2008})
  \bibinfo{pages}{2049--2067}.

\bibitem[{Navot(1998)}]{navot_elastic_1998}
\bibinfo{author}{Y.~Navot}, \bibinfo{title}{Elastic membranes in viscous shear
  flow}, \bibinfo{journal}{Phys. Fluid}
  \bibinfo{volume}{10}~(\bibinfo{number}{8}) (\bibinfo{year}{1998})
  \bibinfo{pages}{1819--1833}.

\bibitem[{Guo et~al.(2002)Guo, Zheng, and Shi}]{guo_discrete_2002}
\bibinfo{author}{Z.~Guo}, \bibinfo{author}{C.~Zheng}, \bibinfo{author}{B.~Shi},
  \bibinfo{title}{Discrete Lattice Effects on the Forcing Term in the Lattice
  Boltzmann Method}, \bibinfo{journal}{Phys. Rev. E}
  \bibinfo{volume}{65}~(\bibinfo{number}{4}) (\bibinfo{year}{2002})
  \bibinfo{pages}{046308}.

\bibitem[{Kr\"uger et~al.(2009)Kr\"uger, Varnik, and Raabe}]{kruger_shear_2009}
\bibinfo{author}{T.~Kr\"uger}, \bibinfo{author}{F.~Varnik},
  \bibinfo{author}{D.~Raabe}, \bibinfo{title}{Shear stress in lattice Boltzmann
  simulations}, \bibinfo{journal}{Phys. Rev. E}
  \bibinfo{volume}{79}~(\bibinfo{number}{4}) (\bibinfo{year}{2009})
  \bibinfo{pages}{046704--14}.

\bibitem[{Frisch et~al.(1986)Frisch, Hasslacher, and
  Pomeau}]{frisch_lattice-gas_1986}
\bibinfo{author}{U.~Frisch}, \bibinfo{author}{B.~Hasslacher},
  \bibinfo{author}{Y.~Pomeau}, \bibinfo{title}{Lattice Gas Automata for the
  Navier-Stokes Equation}, \bibinfo{journal}{Phys. Rev. Lett.}
  \bibinfo{volume}{56}~(\bibinfo{number}{14}) (\bibinfo{year}{1986})
  \bibinfo{pages}{1505}.

\bibitem[{Holdych et~al.(2004)Holdych, Noble, Georgiadis, and
  Buckius}]{holdych_truncation_2004}
\bibinfo{author}{D.~J. Holdych}, \bibinfo{author}{D.~R. Noble},
  \bibinfo{author}{J.~G. Georgiadis}, \bibinfo{author}{R.~O. Buckius},
  \bibinfo{title}{Truncation Error Analysis of Lattice Boltzmann Methods},
  \bibinfo{journal}{J. Comput. Phys.}
  \bibinfo{volume}{193}~(\bibinfo{number}{2}) (\bibinfo{year}{2004})
  \bibinfo{pages}{595--619}.

\bibitem[{Le and Zhang(2009)}]{le_boundary_2009}
\bibinfo{author}{G.~Le}, \bibinfo{author}{J.~Zhang}, \bibinfo{title}{Boundary
  slip from the immersed boundary lattice Boltzmann models},
  \bibinfo{journal}{Phys. Rev. E} \bibinfo{volume}{79}~(\bibinfo{number}{2})
  (\bibinfo{year}{2009}) \bibinfo{pages}{026701--8}.

\bibitem[{Skalak et~al.(1973)Skalak, Tozeren, Zarda, and
  Chien}]{skalak_strain_1973}
\bibinfo{author}{R.~Skalak}, \bibinfo{author}{A.~Tozeren},
  \bibinfo{author}{R.~P. Zarda}, \bibinfo{author}{S.~Chien},
  \bibinfo{title}{Strain Energy Function of Red Blood Cell Membranes},
  \bibinfo{journal}{Biophys. J.} \bibinfo{volume}{13}~(\bibinfo{number}{3})
  (\bibinfo{year}{1973}) \bibinfo{pages}{245--264}.

\bibitem[{Bagchi et~al.(2005)Bagchi, Johnson, and
  Popel}]{bagchi_computational_2005}
\bibinfo{author}{P.~Bagchi}, \bibinfo{author}{P.~C. Johnson},
  \bibinfo{author}{A.~S. Popel}, \bibinfo{title}{Computational Fluid Dynamic
  Simulation of Aggregation of Deformable Cells in a Shear Flow},
  \bibinfo{journal}{J. Biomech. Eng.: T. ASME}
  \bibinfo{volume}{127}~(\bibinfo{number}{7}) (\bibinfo{year}{2005})
  \bibinfo{pages}{1070--1080}.

\bibitem[{Canham(1970)}]{canham_minimum_1970}
\bibinfo{author}{P.~B. Canham}, \bibinfo{title}{The minimum energy of bending
  as a possible explanation of the biconcave shape of the human red blood
  cell}, \bibinfo{journal}{J. Theor. Biol.}
  \bibinfo{volume}{26}~(\bibinfo{number}{1}) (\bibinfo{year}{1970})
  \bibinfo{pages}{61--81}.

\bibitem[{Helfrich(1973)}]{helfrich_elastic_1973}
\bibinfo{author}{W.~Helfrich}, \bibinfo{title}{Elastic properties of lipid
  bilayers: theory and possible experiments}, \bibinfo{journal}{Z. Naturforsch.
  C} \bibinfo{volume}{28}~(\bibinfo{number}{11}) (\bibinfo{year}{1973})
  \bibinfo{pages}{693--703}.

\bibitem[{Svetina and Žekš(1989)}]{svetina_membrane_1989}
\bibinfo{author}{S.~Svetina}, \bibinfo{author}{B.~Žekš},
  \bibinfo{title}{Membrane bending energy and shape determination of
  phospholipid vesicles and red blood cells}, \bibinfo{journal}{Eur. Biophys.
  J.} \bibinfo{volume}{17}~(\bibinfo{number}{2}) (\bibinfo{year}{1989})
  \bibinfo{pages}{101--111}.

\bibitem[{Gompper and Schick(2008)}]{gompper2008soft}
\bibinfo{author}{G.~Gompper}, \bibinfo{author}{M.~Schick}, \bibinfo{title}{Soft
  Matter: Lipid Bilayers and Red Blood Cells}, \bibinfo{publisher}{Wiley-VCH},
  \bibinfo{year}{2008}.

\bibitem[{Charrier et~al.(1989)Charrier, Shrivastava, and
  Wu}]{charrier_free_1989}
\bibinfo{author}{J.~Charrier}, \bibinfo{author}{S.~Shrivastava},
  \bibinfo{author}{R.~Wu}, \bibinfo{title}{Free and constrained inflation of
  elastic membranes in relation to thermoforming –– non-axisymmetric
  problems}, \bibinfo{journal}{J. Strain Anal. Eng.}
  \bibinfo{volume}{24}~(\bibinfo{number}{2}) (\bibinfo{year}{1989})
  \bibinfo{pages}{55--74}.

\bibitem[{Shrivastava and Tang(1993)}]{shrivastava_large_1993}
\bibinfo{author}{S.~Shrivastava}, \bibinfo{author}{J.~Tang},
  \bibinfo{title}{Large deformation finite element analysis of non-linear
  viscoelastic membranes with reference to thermoforming}, \bibinfo{journal}{J.
  Strain Anal. Eng.} \bibinfo{volume}{28}~(\bibinfo{number}{1})
  (\bibinfo{year}{1993}) \bibinfo{pages}{31--51}.

\bibitem[{Yang et~al.(2009)Yang, Zhang, Li, and He}]{yang_smoothing_2009}
\bibinfo{author}{X.~Yang}, \bibinfo{author}{X.~Zhang}, \bibinfo{author}{Z.~L.},
  \bibinfo{author}{G.-W. He}, \bibinfo{title}{A smoothing technique for
  discrete delta functions with application to immersed boundary method in
  moving boundary simulations}, \bibinfo{journal}{J. Comput. Phys.}
  \bibinfo{volume}{228}~(\bibinfo{number}{20}) (\bibinfo{year}{2009})
  \bibinfo{pages}{7821--7836}.

\bibitem[{Doddi and Bagchi(2008)}]{doddi_lateral_2008}
\bibinfo{author}{S.~K. Doddi}, \bibinfo{author}{P.~Bagchi},
  \bibinfo{title}{Lateral migration of a capsule in a plane Poiseuille flow in
  a channel}, \bibinfo{journal}{Int. J. Multiphase Flow}
  \bibinfo{volume}{34}~(\bibinfo{number}{10}) (\bibinfo{year}{2008})
  \bibinfo{pages}{966--986}.

\bibitem[{Wu et~al.(2009)Wu, Shu, and Zhang}]{wu_simulation_2009}
\bibinfo{author}{J.~Wu}, \bibinfo{author}{C.~Shu}, \bibinfo{author}{Y.~H.
  Zhang}, \bibinfo{title}{Simulation of incompressible viscous flows around
  moving objects by a variant of immersed boundary-lattice Boltzmann method},
  \bibinfo{journal}{Int. J. Numer. Meth. Fluids}
  \bibinfo{volume}{62}~(\bibinfo{number}{3}) (\bibinfo{year}{2009})
  \bibinfo{pages}{327--354}.

\bibitem[{Rineau and Yvinec(2008)}]{cgal:ry-smg-08}
\bibinfo{author}{L.~Rineau}, \bibinfo{author}{M.~Yvinec}, \bibinfo{title}{3D
  Surface Mesh Generation}, in: \bibinfo{editor}{C.~E. Board} (Ed.),
  \bibinfo{booktitle}{{CGAL} User and Reference Manual}, \bibinfo{edition}{3.4}
  edn., \bibinfo{note}{\texttt{http://www.cgal.org}}, \bibinfo{year}{2008}.

\bibitem[{Geuzaine and Remacle(2009)}]{Gmsh:website}
\bibinfo{author}{C.~Geuzaine}, \bibinfo{author}{J.-F. Remacle},
  \bibinfo{title}{Gmsh: a three-dimensional finite element mesh generator with
  built-in pre- and post-processing facilities.},
  \bibinfo{note}{\texttt{http://www.geuz.org/gmsh/}}, \bibinfo{year}{2009}.

\bibitem[{Keller and Skalak(1982)}]{keller_motion_1982}
\bibinfo{author}{S.~R. Keller}, \bibinfo{author}{R.~Skalak},
  \bibinfo{title}{Motion of a {Tank-Treading} Ellipsoidal Particle in a Shear
  Flow}, \bibinfo{journal}{J. Fluid Mech.} \bibinfo{volume}{120}
  (\bibinfo{year}{1982}) \bibinfo{pages}{27--47}.

\bibitem[{Schmid-Sch\"onbein and Wells(1969)}]{schmid-schonbein_fluid_1969}
\bibinfo{author}{H.~Schmid-Sch\"onbein}, \bibinfo{author}{R.~Wells},
  \bibinfo{title}{Fluid {Drop-Like} Transition of Erythrocytes under Shear},
  \bibinfo{journal}{Science} \bibinfo{volume}{165}~(\bibinfo{number}{3890})
  (\bibinfo{year}{1969}) \bibinfo{pages}{288--291}.

\bibitem[{He et~al.(1997)He, Zou, Luo, and Dembo}]{he_analytic_1997}
\bibinfo{author}{X.~He}, \bibinfo{author}{Q.~Zou}, \bibinfo{author}{L.-S. Luo},
  \bibinfo{author}{M.~Dembo}, \bibinfo{title}{Analytic Solutions of Simple
  Flows and Analysis of Nonslip Boundary Conditions for the Lattice Boltzmann
  BGK Model}, \bibinfo{journal}{J. Stat. Phys.}
  \bibinfo{volume}{87}~(\bibinfo{number}{1}) (\bibinfo{year}{1997})
  \bibinfo{pages}{115--136}.

\bibitem[{Anderson et~al.(1998)Anderson, McFadden, and
  Wheeler}]{anderson_diffuse-interface_1998}
\bibinfo{author}{D.~M. Anderson}, \bibinfo{author}{G.~B. McFadden},
  \bibinfo{author}{A.~A. Wheeler}, \bibinfo{title}{{Diffuse-Interface} Methods
  in Fluid Mechanics}, \bibinfo{journal}{Ann. Rev. Fluid Mech.}
  \bibinfo{volume}{30} (\bibinfo{year}{1998}) \bibinfo{pages}{139--165}.

\end{thebibliography}

\end{document}